\documentclass[12pt,preprint,useAMS,usenatbib]{emulateapj}
\usepackage{amsmath,amssymb}
\usepackage{graphicx, txfonts}
\usepackage{setspace}
\usepackage{natbib}
\usepackage{ulem}
\usepackage{url}
\usepackage{multirow}

\newcommand{\CII}{C\,\textsc{ii}}

\newcommand{\HII}{\textrm{H}\,\textsc{ii}}

\newcommand{\OI}{O\,\textsc{i}}

\shorttitle{The ancient population of CEMP-no stars}
\shortauthors{Cooke \&\ Madau}

\begin{document}

\title{Carbon-Enhanced Metal-Poor stars: Relics from the Dark Ages}

\author{
Ryan J. Cooke\altaffilmark{1,2} and
Piero Madau\altaffilmark{1}
}

\altaffiltext{1}{Department of Astronomy and Astrophysics, UCO/Lick Observatory, University of California, Santa Cruz, CA 95064, USA}
\altaffiltext{2}{Morrison/Hubble Fellow;~~~~email: rcooke@ucolick.org}
\date{\today}

\begin{abstract}
We use detailed nucleosynthesis calculations and a realistic prescription
for the environment of the first stars to explore the first episodes of
chemical enrichment that occurred during the dark ages.
Based on these calculations, we propose a novel explanation for the
increased prevalence of carbon-enhanced metal-poor (CEMP) stars
with decreasing Fe abundance: The observed chemistry for the most
metal-poor Galactic halo stars is the result of an intimate link
between the explosions of the first stars and their host minihalo's
ability to retain its gas. 
Specifically, high-energy supernovae produce a near solar ratio of C/Fe,
but are effective in evacuating the gas from their host minihalo,
thereby suppressing the formation of a second generation of stars.
On the other hand, minihalos that host low-energy supernovae are
able to retain their gas and form a second stellar generation but, as
a result, the second stars are born with a supersolar ratio of C/Fe.
Our models are able to accurately reproduce the observed
distributions of [C/Fe] and [Fe/H], as well as the fraction of CEMP stars
relative to non-CEMP stars as a function of [Fe/H] without any free
parameters. We propose
that the present lack of chemical evidence for very massive stars
($\gtrsim140~{\rm M}_{\odot}$) that ended their lives as a highly
energetic pair-instability supernova, does not imply that such stars
were rare or did not exist; the chemical products of these very massive
first stars may have been evacuated from their host minihalo, and
were never incorporated into subsequent generations of stars.
Finally, our models suggest that the most Fe-poor stars currently
known may have seen the enrichment from a small multiple of
metal-free stars, and need not have been exclusively enriched
by a solitary first star. These calculations also add further support
to the possibility that some of the surviving dwarf satellite
galaxies of the Milky Way are the relics of the first galaxies.
\end{abstract}


\section{Introduction}

The first episodes of star formation arguably mark one of the most
important chemical transformations of our Universe, a process where
the initially pristine H and He nuclei were converted into heavy elements
for the first time. Although our current understanding of these first stars
is far from complete, the qualitative picture is well-established
(see e.g. \citet{Bro13} for a recent review); the first stars formed in
$\sim10^{6}\,{\rm M}_{\odot}$ dark matter minihalos at redshifts
$15\lesssim z \lesssim 30$ \citep{HaiThoLoe96,Teg97}, and were
predominantly a massive population of stars owing to a lack of
efficient cooling routes in the primordial gas. This picture continues
to receive strong support from cosmological hydrodynamic simulations
\citep[][and other references cited herein]{AbeBryNor02,BroCopLar02}.
Despite the great success of these foundations,
some of the most fundamental properties of the first
stars have yet to be pinned down, including the distribution
of their stellar mass \citep{Gre11,StaBro13,Hir14},
the mixing properties and energy released during their putative
supernova (SN) explosion \citep{TomUmeNom07,HegWoo10,LimChi12},
their binary fraction \citep{StaBro13},
the distribution of their host minihalo masses \citep{Yos03,OShNor07,Hir14},
and the finer details of their chemical enrichment
\citep{MadFerRee01,BroYosHer03,ScaSchFer03,Tum06,WisAbe08,Kom10,Mai11,Rit12}.

If metal-free stars (or gas of primordial composition) still exist in the present day
Universe, they have not yet been found. At present, the only primordial
environments that have been uncovered are at high redshift
\citep{FumOMePro11,Sim12}; the discovery of such systems
is crucial for providing a direct physical insight into the formation environments
of the first stars. Otherwise, our complete understanding of the first stars relies
on simulated universes and other indirect probes, such as the chemical
yields from metal-free stars. Fortunately, there now exist a small handful of
gas-rich systems at high redshift with metallicities $\lesssim1/1000$
of solar that might be solely enriched by the earliest stellar populations
\citep{Pet08,Ell10,Coo11a,Coo11b,CooPetMur12}. These recent
advances at high redshift offer a new and highly complementary approach
for studying the nucleosynthesis from the first stars, which has traditionally been
pursued by studying the most metal-poor, second generation stars in the
local Universe \citep{Cay04,Lai08,Roe14}.

Perhaps one of the most striking results that has surfaced from studies of metal-poor
Galactic halo stars is the great diversity in their chemical composition.
Specifically, around one-quarter of all stars with [Fe/H]\footnote{Throughout this
paper, we adopt the standard notation
[X/Y]$\equiv\log\,N({\rm X})/N({\rm Y}) - \log\,({\rm X}/{\rm Y})_{\odot}$,
where $N({\rm X})/N({\rm Y})$ is the number abundance ratio of element X
relative to element Y, and the ${\odot}$ symbol refers to the solar value, taken
from \citet{Asp09}.}~$<-2.0$ exhibit a
stark underabundance of heavy elements (e.g. Fe) relative to lighter
metals (e.g. C) \citep{BeeChr05,Luc06,Aok07,Nor13,Yon13}, which becomes
even more pronounced at lower [Fe/H] abundance. Such stars are collectively
known as Carbon-Enhanced Metal-Poor (CEMP)\footnote{The `CEMP' label
is sometimes considered a misnomer, since \textit{some} of the CEMP stars
are believed to be underabundant in heavy elements, rather than enhanced
in the low atomic number elements.} stars, and are somewhat arbitrarily defined
to have [C/Fe]$\,\,>+0.7$ \citep[see e.g.][]{Aok07}.

Large samples of CEMP stars have shown that there appears to be
four distinct sub-classes, which are usually defined on the basis of
their neutron-capture elements (see \citealt{BeeChr05} for further details):
(1) and (2) CEMP-$s$ and CEMP-$r$ stars, which are
enhanced in the $s$-process and $r$-process relative to the solar
composition, respectively;
(3) CEMP-$rs$ stars, which exhibit enhancements in both the $r$- and $s$-
neutron-capture processes; and
(4) CEMP-no stars, with a normal abundance pattern
for their neutron-capture elements.

Several mechanisms have been proposed to explain the origin of
these distinct classes of CEMP stars. For example, the CEMP-$s$
stars are known to be members of a binary system where an
asymptotic giant branch (AGB) star transferred carbon-rich
(and $s$-process-rich) material onto a low-mass long-lived
companion star that we observe today to be anomalously
rich in C \citep[e.g.][]{McC85}. Indeed, long-term monitoring of the
radial velocities of such stars seems to confirm this picture \citep{Luc05,Coh06,Sta14}.
The origin of the CEMP-no stars, however, is still a matter of
debate. While some CEMP-no stars could be an extension
of the CEMP-$s$ class to lower metallicity and lower
$s$-process enhancement \citep{CamLat08,Mas10,KarLat14},
there are several lines of evidence that support an alternative
origin for the carbon enhancement in some CEMP-no stars.
First, the most Fe-poor CEMP-no stars 
exhibit the largest C/Fe ratios \citep{Chr02,Fre05,Nor07,Kel14}
and appear to be more common at lower metallicities
\citep{BeeChr05,Aok07,Car12,Yon13,Car14}.
In addition, the overabundance of each element
relative to Fe decreases with increasing atomic number.
This evidence suggests in fact that \textit{iron} is
strongly depressed in these stars, rather than carbon being strongly
enhanced (the latter is the case for CEMP-$s$ stars). Furthermore, the
binarity properties of CEMP-no stars appear to be markedly different
from the CEMP-$s$ class \citep{HanAndNor13,Nor13}. Collectively,
these observations suggest that most of the CEMP-no stars
were probably born out of gas enriched by massive stars
that ended their lives as Type II SNe with low levels of mixing
and a high degree of fallback \citep{UmeNom03,Rya05,Ish14}.

However, there remains one aspect of this model that has yet
to receive a satisfactory explanation --- why does the fraction
of CEMP-no stars become more common relative to non-CEMP
stars at lower Fe abundance? Is it possible to observe a star
that has [Fe/H]~$\simeq-5.0$ and [C/Fe]~$\simeq0.0$?
One possible solution to explain why no such star has been
observed was proposed by
\citet[][see also \citet{FreJohBro07}]{BroLoe03}, who require
a minimum level of [C/H] and [O/H] to be present
in the birth cloud of the second generation stars to allow efficient
cooling by the fine-structure transitions of \CII\ and \OI.
In this picture, the cooling by \CII\ and \OI\ fine-structure
lines allows the gas to fragment to smaller mass scales
and form lower mass (hence longer lived) stars that
have lived until the present day when we observe them.
This model therefore implies that the increased
fraction of CEMP-no stars at low metallicity is a selection effect,
since we can only observe the lowest mass, second generation
stars born out of gas with a minimum abundance of C and O,
irrespective of the [Fe/H] abundance.
This model was called into question with the discovery of the
Leo star (\citealt{Caf11}; [Fe/H]$\,=-4.73$, [C/Fe]$\,\le+0.93$),
which is the most metal-deficient star currently known. To explain
this almost primordial star, several authors have appealed to
models of dust-induced fragmentation
\citep{KleGloCla12,Sch12,JiFreBro14}, which is able to
assist low-mass star formation in the absence of high levels
of C and O. In particular, \citet{JiFreBro14} propose that both
fine-structure cooling and fragmentation by dust are required
in order to explain the population of CEMP-no stars in addition
to the Leo star.

In this paper, we present a simple solution to explain why
carbon enhancements are more common at the lowest Fe
abundance. Our interpretation is complementary to the methods
described above, but does not rely on the detailed chemistry
of the gas giving rise to the second generation of stars. We
propose that the increased fraction of CEMP-no stars at lower
metallicity is a selection effect due to an intimate link between
the energy released by the SNe of the first stars and
the host minihalo's ability to retain its gas: Qualitatively, a
Type II SN with a high explosion energy is able to
remove more Fe from the core of the massive star, and
tends to produce a solar relative abundance of C to Fe
(i.e. [C/Fe]$\,\sim+0.0$). However, the high energy of such
an explosion is more likely to evacuate the gas from the host
minihalo thereby inhibiting the formation of second generation
non-CEMP stars. On the other hand, low-energy SNe
experience larger degrees of fallback which locks up the Fe
in the compact remnant of the massive star, while still yielding a
full complement of C (i.e. [C/Fe]$\,\gg+0.0$). In this case, the low
energy that is released by the SN is unable to evacuate
the gas from the host potential, thereby allowing a generation of
CEMP stars to form in the same minihalo, given sufficient time
for the gas to cool.

In the following section, we provide the details of our chemical
enrichment models for the minihalos that hosted the first stars.
We present the main findings of our work in Section~\ref{sec:results},
where we explore the mass distribution of the halos that are able
to retain their gas, the chemical properties of the second stars,
and the fraction of CEMP stars that are born in the first galaxies.
We present our main conclusions in
Section~\ref{sec:conc}. Throughout, we have assumed
a \textit{Planck} cosmology \citep{Efs13}, with
$H_{0}=67.3~{\rm km}~{\rm s}^{-1}~{\rm Mpc}^{-1}$,
$\Omega_{\rm M} = 0.315$ and
$\Omega_{\rm \Lambda} = 0.685$.

\section{Model Simulations}
\label{sec:sim}

The principal goal of this paper is to derive the metallicity
and chemical abundance patterns for the second generation
of stars in a cosmological context. In this section, we outline
our model calculations that consider the full evolution
of the first stars, and their impact on the gas confined by the
host dark matter minihalos.

\subsection{Host Minihalo Properties}

Within the standard cold dark matter paradigm, detailed calculations
of structure formation have shown that the first stars typically formed
in minihalos of mass $\sim10^6~{\rm M}_{\odot}$ at redshifts $15\lesssim z \lesssim30$
\citep{Teg97,AbeBryNor02,BroCopLar02}. This general picture has
been confirmed by more recent work, where the minimum minihalo mass
that is able to produce sufficient molecular hydrogen and therefore cool
efficiently to successfully host a first generation of stars is
$\sim2\times10^{5} {\rm M}_{\odot}$ (e.g. \citealt{Yos03,OShNor07}). We therefore
start our calculations by drawing a random redshift and minihalo virial mass ($M_{\rm 200}$)
from the halo mass function derived by \citet{Reed07}\footnote{The halo mass
functions were generated using \textsc{HMFcalc} \citep{MurPowRob13}.},
with a lower-mass cut-off at $\sim2\times10^{5}~{\rm M}_{\odot}$ in the redshift
interval $15-30$.
The halo concentration, $c_{200}$, is chosen by extrapolating the halo-mass--concentration
relation derived by \citet{Pra12}, and is typically in the range $4-7$ for the
halo masses and redshift range that we consider here.  Prior to the formation of the first stars, we
assume that the gas confined to the dark matter minihalo obeys an isothermal
density profile, which is motivated by several cosmological simulations
\citep[e.g.][]{AlvBroSha06,OShNor07,Wha08}. We assume the initial
temperature of the gas is set by the virial temperature of the minihalo scaled
(by a factor of $0.6$) to match cosmological simulations (see e.g. Figure 19
of \citealt{OShNor07}), and is of the form:
\begin{equation}
T_{\rm gas}(0) = 0.6\times100^{1/3}\,\frac{\mu_{0} m_{\rm H}}{2\,k_{\rm B}}\,(G\,M_{\rm 200}\,H(z_{\rm coll}))^{2/3},
\end{equation}
where $H(z_{\rm coll})$ is the Hubble parameter at the minihalo collapse redshift
($z_{\rm coll}$)\footnote{$z_{\rm coll}$ is defined to be the point at which the
host minihalo's gas reaches high enough densities for the first stars to form -- see \citet{OShNor07}.},
and $m_{\rm H}$, $k_{\rm B}$, and $G$ are the proton mass, Boltzmann constant
and Newton gravitational constant respectively. The mean molecular weight
in the initially neutral primordial gas is $\mu_{0}\simeq1.23$.

\subsection{The First Stars}
\label{sec:stars}

\begin{figure*}
  \centering
  {\includegraphics[angle=0,width=8.5cm]{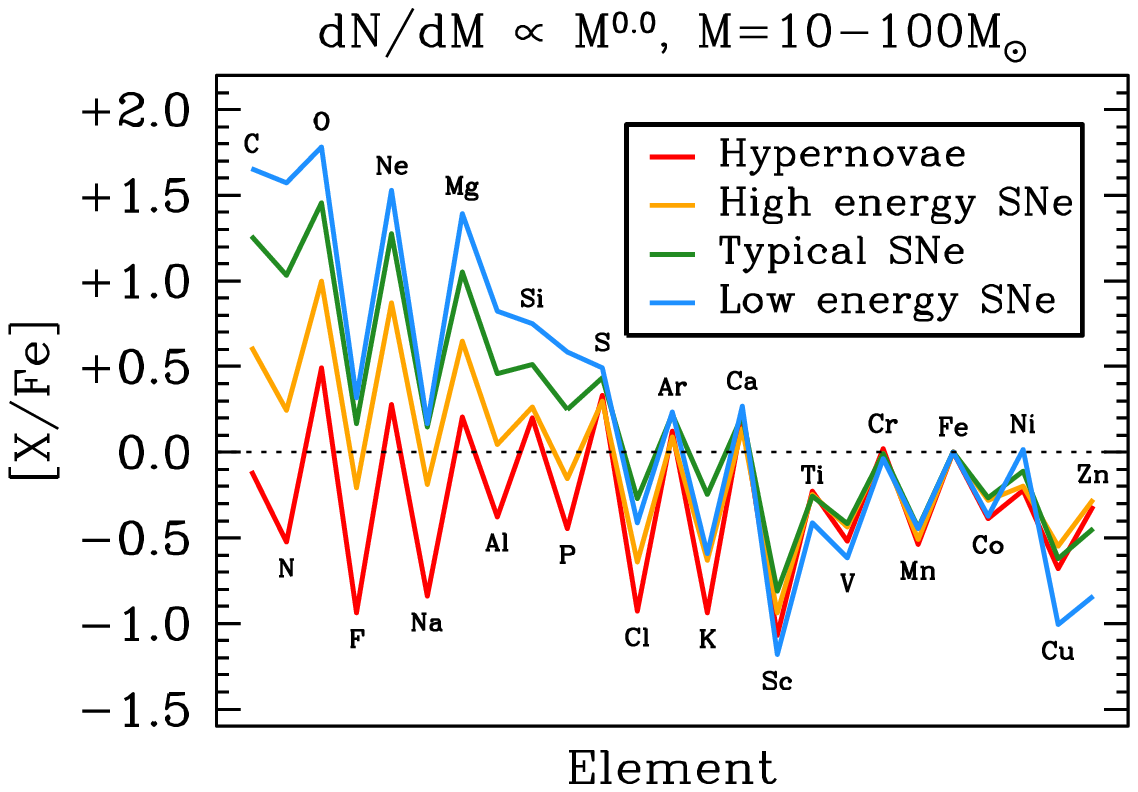}\hspace{0.5cm}
  \includegraphics[angle=0,width=8.5cm]{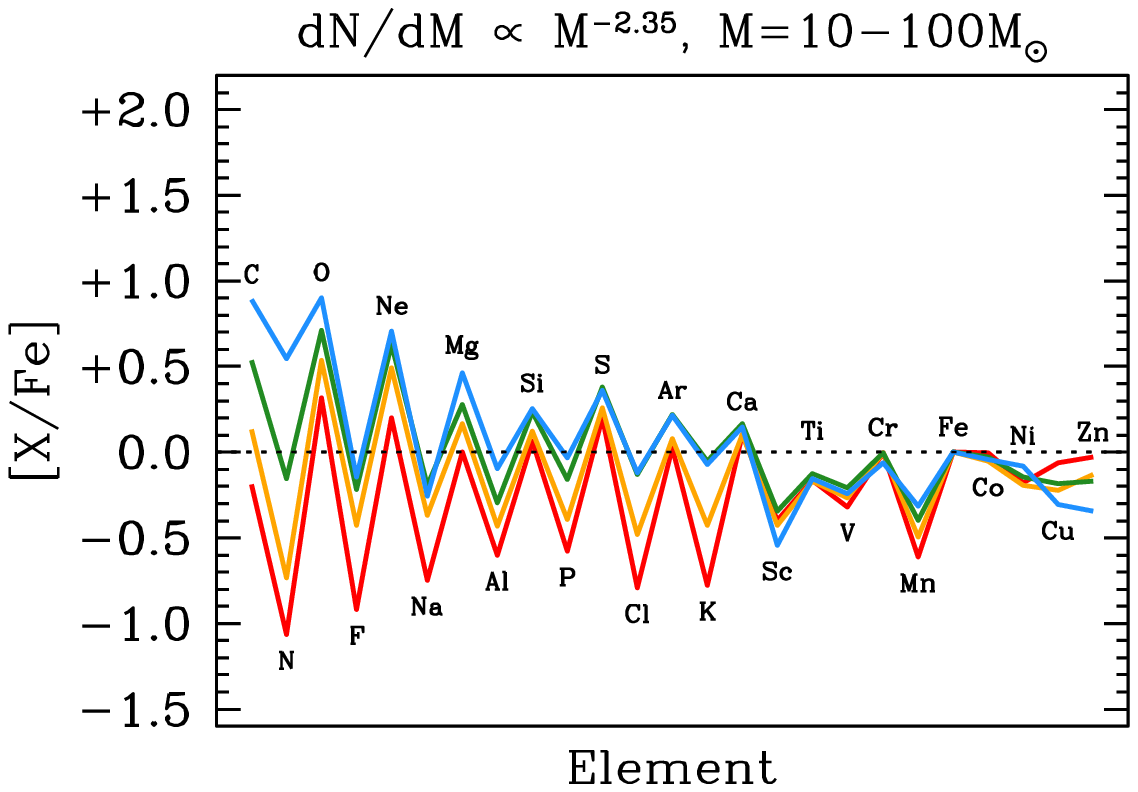}}\\
  \caption{ 
  The IMF-weighted chemical yields for zero-metallicity stars with masses in
  the range $10-100\,{\rm M}_{\odot}$ \citep{HegWoo10} are integrated over a
  flat IMF (left panel) and a Salpeter-like power-law IMF (right panel). For
  each IMF, we color-code the yields by the energy released during the
  explosion (see the text). This figure demonstrates that the highest mass massive stars
  coupled with low-energy explosions can produce a strong enhancement
  in the C/Fe ratio, as noted previously \citep[e.g.][]{UmeNom03}.
 }
  \label{fig:yieldcalc}
\end{figure*}

In our study, we investigate zero-metallicity massive stars in the mass
range $10-100\,{\rm M}_{\odot}$ that would have ended their lives as
Type-II core-collapse SN. In principle, the mass function
of the first stars might extend to much higher masses, which we
do not consider here for the following reasons: Stars more massive than
$\sim100\,{\rm M}_{\odot}$ exhibit a pulsational instability due to the creation
of electron/positron pairs in abundance \citep{Led41}. Stars in the mass
range $100-140\,{\rm M}_{\odot}$ experience a series of such pulsations,
causing the star to eject the outermost layers before it quietly collapses to a
black hole \citep{WooBliHeg07}. Stellar evolution models of such stars
have indicated that radiative mass loss provides negligible metal enrichment
\citep{MarChiKud03}. However, the pulsational winds from such stars might play a role in
enriching early galaxies with CNO, while producing no trace of
the Fe-peak elements (\citealt{Che14a,WooHeg14}; but see \citealt{BarHegWoo01}).
In this context, detailed yield calculations for pulsational pair-instability SNe
are highly desirable (see the discussion in Section~\ref{sec:nucleo}).
For stars in the mass range $\sim140-260\,{\rm M}_{\odot}$, the first
instance of the pair-instability is sufficient to entirely disrupt the
star \citep{HegWoo02,Che14b}, leading to a pair-instability supernova (PISN)
with a typical explosion energy of $\gtrsim10^{52}$~erg\footnote{We
also note that there is a narrow range of stellar mass between
$140-150~{\rm M}_{\odot}$ where PISNe with explosion
energies as low as $5\times10^{51}$~erg could exist.}; such a high explosion
energy is sufficient to evacuate the host minihalo of all its gas, thereby
prohibiting the formation of a second generation of stars in this minihalo
\citep{Gre07,Wha08} --- evacuating the gas reservoir defeats the
primary goal of our study to investigate the halos that host the
second generation of stars. Stars with mass $\gtrsim260\,{\rm M}_{\odot}$
are sufficiently massive that the onset of the pair-instability and the
subsequent contraction and burning is unable to halt the gravitational
collapse of the star, thus providing no metal enrichment \citep{FryWooHeg01}.
In summary, metal-free stars in the mass range $10-100\,{\rm M}_{\odot}$
are currently considered to be the dominant source of metal enrichment
for the second stars, with perhaps an additional contribution from the
pulsational pair-instability SNe.

We have thus adopted the model yield calculations performed by
\citet{HegWoo10}, which provide illustrative yields for 120
different stellar masses in the range $10-100\,{\rm M}_{\odot}$ (with
some models differing in mass by just $0.1\,{\rm M}_{\odot}$). Each
star was exploded with a piston located at the base of the
oxygen shell (where the entropy per baryon $\sim4\,k_{\rm B}$),
providing a final kinetic energy to the ejecta at infinity of either
0.3, 0.6, 0.9, 1.2, 1.5, 1.8, 2.4, 3, 5, or 10
in units of $10^{51}~{\rm erg}$. A simple prescription of
the mixing between the stellar layers is implemented
by a moving boxcar filter of a set width (14 different
widths are considered here).
Therefore, the final set of yields for the stars considered
in our study totals 16,800 models. After drawing a stellar
mass from the appropriate initial mass function (IMF; see below),
we randomly draw an explosion energy and level of mixing from
a uniform distribution. Given that we have no handle on the mixing
and explosion properties of the first stars, our assumption
allows sufficient flexibility to explore all of the available
parameter space. In Figure~\ref{fig:yieldcalc}, we present
the IMF-weighted yields for illustration purposes, where
we separately plot the integrated yields for SNe that release
low energy ($0.3-0.6\times10^{51}\,{\rm erg}$; blue curves),
typical energies ($0.9-1.5\times10^{51}\,{\rm erg}$; green curves),
high energies ($1.8-3\times10^{51}\,{\rm erg}$; orange curves), and
hypernovae ($5-10\times10^{51}\,{\rm erg}$; red curves).

The current generation of cosmological hydrodynamic simulations
that follow the collapse and formation of the first stars suggest that
either a single, binary, or small multiple of metal-free stars form in
the center of dark matter minihalos \citep{TurAbeOsh09,StaGreBro10,StaBro13,Hir14}.
We therefore assign each minihalo a single metal-free star (this assumption is relaxed
in Section~\ref{sec:multiples}, where we allow small multiples of metal-free
stars to form in a given dark matter minihalo), drawn from a flat stellar IMF,
i.e. ${\rm d}N/{\rm d}m_{\star}\propto m_{\star}^{0}$, in the mass
range $10-100\,{\rm M}_{\odot}$. This assumption is supported by several recent
numerical simulations \citep{StaBro13,Hir14}. However, the IMF of the first stars
is still highly uncertain, and might instead represent a flat distribution in
${\rm d}N/{\rm d}\log m_{\star}$ \citep{Gre11}. Thus, to demonstrate how
our choice of a flat IMF (in ${\rm d}N/{\rm d}m_{\star}$) influences the results,
we also consider the `extreme' case of a Salpeter-like IMF for the first stars,
in the mass range $10-100\,{\rm M}_{\odot}$. For reference, the characteristic
stellar masses for our chosen IMFs are $55\,{\rm M}_{\odot}$ (flat IMF)
and $22.3\,{\rm M}_{\odot}$ (Salpeter-like IMF).

We also note that the \textit{initial} mass function of the first stars could be very different
from the mass function of stars that provide metal enrichment; there may be several windows
in stellar mass where massive stars collapse directly to a black hole and eject no
metals \citep{WooHeg14}. If such a scenario is realized, the mass function of the
first stars that eject metals may resemble a broken picket fence, rather than a power law.
For this study, we assume that every star born in the mass range $10-100\,{\rm M}_{\odot}$
provides metal enrichment. 

\subsection{Presupernova {\rm H} {\sc ii} Region Properties}

The first stars can have a profound impact on their environment
during their lives. They emit copious quantities of H-ionizing photons
which serve to ionize and thereby heat the surrounding gas to
$\sim10^{4}$ K. After the first stars turn on, the initial (isothermal) density
distribution characterized by the temperature $T_{\rm gas}(0)$ becomes
overpressured and causes the gas to expand. The expansion is bounded
by an isothermal shock, known as a ``champagne'' flow, and is characterised
by the self-similar solutions presented by \citet{Shu02}. The subsequent
time-evolution of the radial gas density distribution for a self-gravitating
isothermal cloud of gas is given by
\begin{equation}
\label{eq:densshu}
\rho(r,t)=\frac{k_{\rm m}\alpha(x)}{4 \pi G t^2}\qquad x=\frac{r}{c_{\rm s}(0)\,t}
\end{equation}
where $c_{s}(0)=k_{\rm B}\,T_{\rm gas}(0)/\mu_{0} m_{\rm H}$ is the
sound speed of the gas before the first stars are born. The constant
$k_{\rm m}=M_{\rm gas}/0.6\,M_{\rm 200}\simeq0.17$ is chosen so
that the total mass of baryons inside the virial radius matches that
observed in cosmological hydrodynamic simulations
\citep[see e.g. Table 2 from][]{OShNor07}. After the first
stars are born, they heat the center of the cloud to a new temperature,
$T_{\rm gas}(t)$, with a corresponding sound speed
$c_{s}(t)=k_{\rm B}\,T_{\rm gas}(t)/\mu_{\rm ion} m_{\rm H}$, where
$\mu_{\rm ion}=0.59$ is the mean molecular weight for fully ionized primordial
gas. By numerically integrating the \citet{Shu02} equations, the reduced
central density, $\alpha(x=0)$, in Equation \ref{eq:densshu} can be directly
related to the ratio of the initial and final sound speeds of the gas,
$\epsilon=c_{s}(0)^2/c_{s}(t)^2$. To minimize computational
time, we generate a fine grid of $\alpha(0)$ values, numerically
integrate the \citet{Shu02} equations and compute the
corresponding values for $\epsilon$. Then, given the initial
and final sound speeds of the gas in each minihalo, we interpolate
the $\alpha(x)$ values.

The boost in thermal energy that is given to the gas by a star during
its life can be estimated by equating the heating rate to the cooling rate
of a pure H gas (i.e. assuming thermal equilibrium in a smooth medium).
The cooling rate that we use includes a contribution from recombination
cooling as well as collisional excitation and ionization cooling
(see \citealt{Cen92} for the relevant fitting formulae that we have
adopted herein). Assuming that recombinations balance photoionizations,
the photoelectron heating rate is then simply
\begin{equation}
\frac{\Gamma_{\rm pe}}{n_{\rm H}\,n_{e}} = \alpha_{\rm B}\,\psi k_{\rm B} T_{\rm eff}
\end{equation}
where $\alpha_{\rm B}$ is the case B recombination coefficient for H,
$\psi$ is the mean photoelectron energy, and $T_{\rm eff}$ is the
effective temperature of the star. To calculate $T_{\rm eff}$ for our
full range of stellar masses, we have generated the following fitting formula for
a metal-free star with mass $m_{\star}$ based on the detailed computations
by \citet{Sch02}:
\begin{equation}
T_{\rm eff} = 6600\,{\rm K}\,\, \times (m_{\star}/{\rm M}_{\odot})^{1.09 - 0.257\log_{10}(m_{\star}/{\rm M}_{\odot})}
\end{equation}
The mean photoelectron energy can then be calculated by
assuming that the first stars radiated a blackbody spectrum,
$B_{\nu}(T_{\rm eff})$, with color temperature $T_{\rm eff}$
\begin{equation}
\label{eq:phe}
\psi k_{\rm B} T_{\rm eff} = \frac{\int_{\nu_{0}}^{\infty}[B_{\nu}(T_{\rm eff})/h\nu]\sigma_{\nu}({\rm H})(h\nu-h\nu_{0})\,{\rm d}\nu}{\int_{\nu_{0}}^{\infty}[B_{\nu}(T_{\rm eff})/h\nu]\sigma_{\nu}({\rm H})\,{\rm d}\nu}
\end{equation}
where $\sigma_{\nu}({\rm H})$ is the photoionization cross-section
for H, and $h\nu_{0}=13.6\,{\rm eV}$ is the photoionization energy
of H.
\begin{figure}
  \centering
  \includegraphics[angle=0,width=8.5cm]{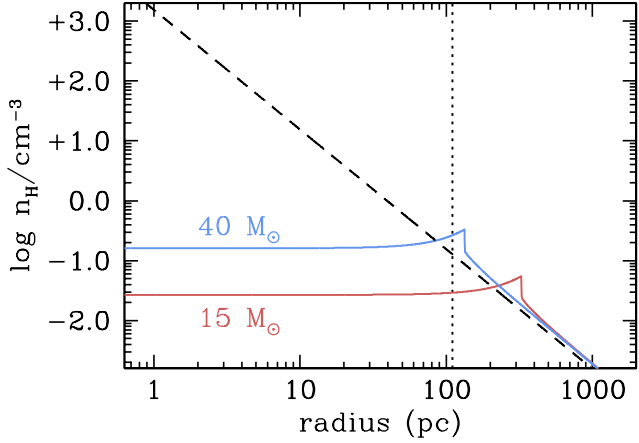}
  \caption{ 
The gas density profile for a $M_{200}=7\times10^{5}\,{\rm M}_{\odot}$
minihalo that collapsed at redshift $z=24$ is shown as the dashed line.
The blue and red lines show the gas density
profile of this minihalo after being heated for the duration of a
$40\,{\rm M}_{\odot}$ and $15\,{\rm M}_{\odot}$ metal-free star's life, respectively.
Note the striking similarity between our model calculations
and that presented in Figures~1 and 2 of \citet[][see also \citealt{AbeWisBry07,Yos07}]{Wha08}.
The virial radius of this minihalo is marked by the vertical
dotted line.
 }
  \label{fig:densprof}
\end{figure}

The gas density profile at the end of the star's life is
defined by Equation~\ref{eq:densshu}, where the time variable
corresponds to the main sequence lifetime of a
Population III star. To estimate the main-sequence lifetime
for the range of stellar masses we consider here, we have
generated the following fitting formula, which is derived from
the computations by \citet{Sch02}:
\begin{equation}
t_{\rm ms} = 1300\,{\rm Myr}\,\, \times (m_{\star}/{\rm M}_{\odot})^{-2.407 + 0.5323\log_{10}(m_{\star}/{\rm M}_{\odot})}
\end{equation}
An example density profile for a $15$ and $40\,{\rm M}_{\odot}$ star occupying
a $7\times10^{5}\,{\rm M}_{\odot}$ minihalo that collapsed at redshift $z=24$ is shown
in Figure~\ref{fig:densprof}. The choice of these parameters allow for a direct comparison with
Figures~1 and 2 from \citet{Wha08}. Although we have not modeled the presupernova
evolution using computationally expensive full three-dimensional (3D) cosmological hydrodynamic
simulations, the striking agreement between our calculations and other, more expensive
simulations \citep[e.g.][]{AbeWisBry07,Yos07,Wha08} demonstrates that we
have captured the relevant physics of the presupernova evolution for the first
stars, and the feedback they imparted on the surrounding gas.
Thus, the computational simplicity of our models allows us to
explore vast regions of parameter space in a fraction of the time.

\subsection{Blastwave Evolution}
\label{sec:blastwave}

The final stage of our modeling procedure accounts for the
Type II core-collapse SN blastwave at the end of a star's life.
\citet{Gre07} have shown that the blastwave evolution for the
first stars is well-approximated by an energy and momentum
conserving formalism. When including the host potential,
however, this calculation becomes computationally expensive
\citep{OstMcK88}\footnote{We note that in the work of \citet{Gre07},
the host potential could be safely ignored, since the kinetic energy
released during the explosion was much larger than the binding
energy of the halo.}. Given that the binding energy of the host minihalo
is comparable to the energy released for some of our adopted
models (see Section~\ref{sec:stars}), we have instead calculated
the stall radius, $r_{\rm stall}$, which corresponds to the radius
when the shock velocity is comparable to local sound speed of
the gas. Specifically,
\begin{equation}
\frac{E_{\rm sh}(r_{\rm stall})}{M_{\rm sw}(r_{\rm stall})} = \frac{k_{\rm B}\,T_{\rm gas}(r_{\rm stall})}{\mu(r_{\rm stall})\,m_{\rm H}}
\end{equation}
where the kinetic energy of the shock in the thin shell
approximation is given by
\begin{equation}
E_{\rm sh}(r_{\rm stall}) = (1-f_{\rm r})\,E_{\rm exp} - \Delta W/2.
\end{equation}
Here
\begin{equation}
\Delta W = \!\!\!\int\limits_{0}^{r_{\rm stall}}\!\!\!G\,[M_{\rm dm}(<r)+M_{\rm sw}(r)]\bigg(\frac{M_{\rm sw}(r)}{r^{2}}-4 \pi \rho(r,t_{\rm ms})\bigg)\,{\rm d}r\quad
\end{equation}
is the change in the total gravitational potential produced by the
blastwave, and $f_{\rm r}=0.7$ is the fraction of the kinetic energy
released by the core-collapse SN explosion ($E_{\rm exp}$) that
is radiated away in the Sedov-Taylor phase \citep{OstMcK88}.
The dark matter mass interior to a radius $r$ (assumed to be static) and
the swept up gas mass are respectively given by:
\begin{eqnarray}
M_{\rm dm}(<r) &=& M_{200}\,\frac{\ln(1+u) - u/(1+u)}{\ln(1+c_{200}) - c_{200}/(1+c_{200})}
\\
M_{\rm sw}(r) &=& 4\pi\int_{0}^{r}\rho(R,t_{\rm ms})\,R^2\,{\rm d}R
\end{eqnarray}
where $u=r\,c_{200}/r_{200}$ and $r_{200}$ is the virial radius.

\subsection{Gas Retention and Metallicity}
\label{sec:retention}
The metallicity of the enriched gas depends on how efficiently
the metals are mixed with the pristine material swept up by the
blastwave. For relatively high-mass minihalos ($\sim10^8~{\rm M}_{\odot}$),
the current state-of-the-art simulations that include a prescription
for metal diffusion indicate that the gas is efficiently mixed within
$\sim300~{\rm Myr}$ \citep{Gre10}. For minihalos of lower mass,
efficient mixing takes place within several tens of Myr \citep{BlaSutKar11,Rit12}.
For now, we assume that the ejected metals are efficiently
mixed with the swept up pristine material interior to the stall
radius, such that the gas will be enriched to a metal abundance:
\begin{equation}
{\rm [Z/H]} = \log_{10}\bigg(\frac{m_{\rm H}}{m_{\rm Z}}\frac{M_{\rm Z}}{X\,M_{\rm sw}(r_{\rm stall})}\bigg) - \log_{10}({\rm Z/H})_{\odot}
\end{equation}
where $m_{\rm H}$ and $m_{\rm Z}$ are the atomic masses of hydrogen
and element `Z' respectively, $M_{\rm Z}$ refers to the total ejected mass of
element Z from the metal-free star and $X=0.75$ is the primordial mass
fraction of H. The solar values for all elements, marked by the $\odot$
symbol, are taken from \citet{Asp09}. We further consider the possibility
of inefficient mixing in Section~\ref{sec:nucleo}.

In order to form a second generation of stars encoded with the chemical
signature of the first stellar generation, we require that the blastwave does
not evacuate the gas from the host dark matter minihalo. Specifically, we assume
that all minihalos where the blastwave lifts the gas to beyond the virial radius will
not form a second generation of stars before merging into more massive parent
systems. Thus, the only halos that will form a second generation of stars are
those where $r_{\rm stall}<r_{200}$. Although the retention or evacuation of gas
from a minihalo can only be correctly treated within a 3D setting \citep{Rit12,Smi14,Rit14},
our prescription allows us to explore large areas of parameter space
that would otherwise be impossible with full 3D hydrodynamic
simulations. In any case, we can test the impact of this assumption,
by considering a stricter limit where the only minihalos that retain
their gas are those where the blastwave does not lift the gas beyond
half the virial radius (see Section~\ref{sec:nucleo}).

\subsection{Population III Multiples}
\label{sec:multiples}

Recent simulations suggest that a given minihalo may form a
binary or small multiple of massive metal-free stars
\citep{TurAbeOsh09,StaGreBro10,Gre12,StaBro13,Hir14},
which could reduce
the chance of gas retention relative to the single star scenario
(K.-J. Chen et al. in preparation).
To simulate the chemical enrichment from a small multiple
of Population III stars in a single dark matter minihalo, we draw
one, two, three, or four stars from the stellar IMF under consideration,
and assume that all stars form at a redshift $z_{\rm coll}$
(i.e. the star formation occurs in a ``burst''). The temperature
of the \HII\ region heated by this small multiple
is calculated as described above, but replacing
$B_{\nu}(T_{\rm eff})$ in Equation~\ref{eq:phe}
by the sum over every star in the minihalo,
$\Sigma_{i}\,B_{\nu}(T_{\rm eff, i})$. The blastwave is
initiated after the maximum stellar lifetime has passed
(i.e. the least massive star that exploded). For multiple stars,
the blastwave is followed as an `equivalent' single explosion
as described in Section~\ref{sec:blastwave}; specifically, the
total energy of the blastwave is equal to the sum of the
energy released from the stellar multiple. This assumption
is the limiting case where it is most difficult for a minihalo
to retain its gas. Similarly, each individual element yield is
totaled for all stars in a minihalo, so that we track the
\textit{total} yield of, for example, carbon from the stellar
population.

Thus, our calculation ignores the physics of SN
kicks that might occur after the most massive star in a
binary explodes \citep{ConKra12}. We note that if Population III
runaways are efficiently kicked from the center of the dark matter
potential, they will produce off-centered explosions, which
provide an enhanced gas retention \citep{BlaSutKar11}.

\subsection{Model Statistics}

For the results of this paper, we are concerned with the halos that
are able to retain their gas after forming the first generation of stars.
To provide reliable statistics, we continuously simulate minihalos for a
given IMF and multiple of stars until we have uncovered at least
$10,000$ models that retain their gas.

\section{Results}
\label{sec:results}

In this section, we present the results from our chemical enrichment
models of the first stars. As expected, the results of our simulations
depend strongly on the energy released by the SNe.
To demonstrate the effect of the explosion energy,
we have divided the models into four bins depending on the
average energy of the SN explosion in a minihalo; the intervals
that we have selected are based on the grid of kinetic energy used in the
nucleosynthesis models \citep[][see Section~\ref{sec:stars}]{HegWoo10}.
Specifically, the blue curves in Figures~\ref{fig:massfunc}--\ref{fig:cempfrac}
show low-energy SNe ($0.3-0.75\times10^{51}\,{\rm erg}$),
the green curves show typical SNe ($0.75-1.65\times10^{51}\,{\rm erg}$),
the orange curves are for high-energy SNe ($1.65-4\times10^{51}\,{\rm erg}$)
and the red curves represent the hypernovae ($4-10\times10^{51}\,{\rm erg}$).
For a given SN energy, we have assumed that all values of
the stellar mixing are equally probable.

\subsection{Halo Mass Distribution of the First Galaxies}
\label{sec:litsyst}

\begin{figure*}
  \centering
  {\includegraphics[angle=0,width=8.0cm]{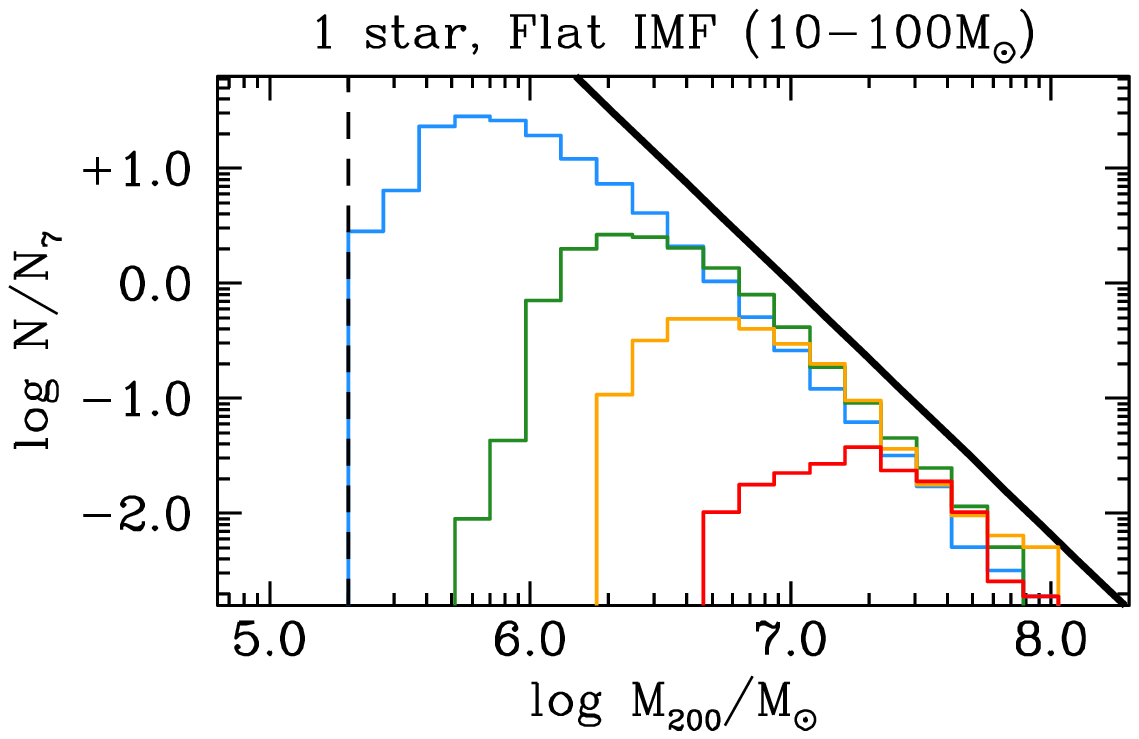}\hspace{0.5cm}
  \includegraphics[angle=0,width=8.0cm]{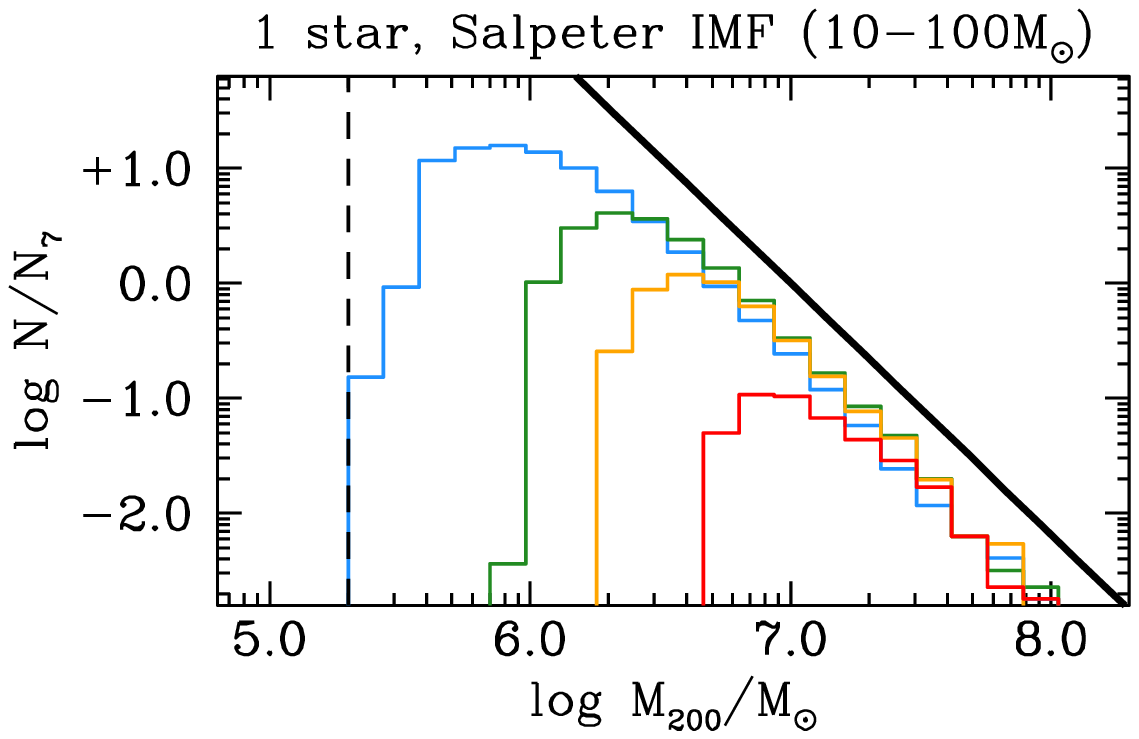}}\\
  \vspace{0.4cm}
  {\includegraphics[angle=0,width=8.0cm]{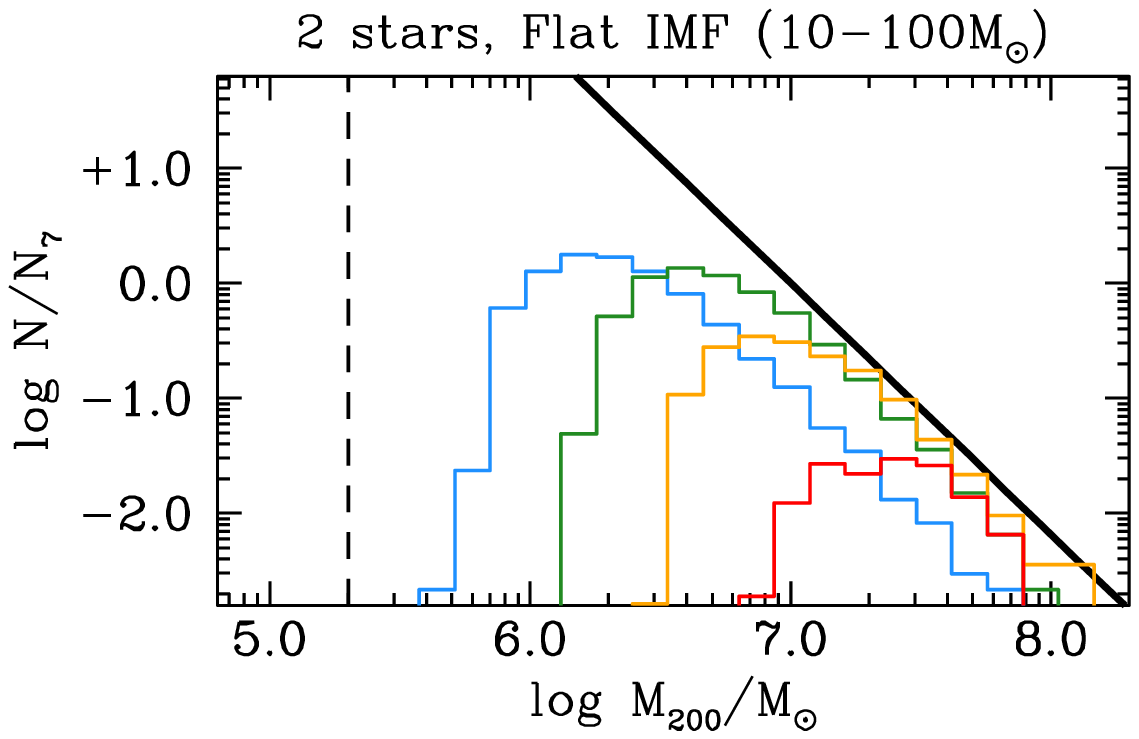}\hspace{0.5cm}
  \includegraphics[angle=0,width=8.0cm]{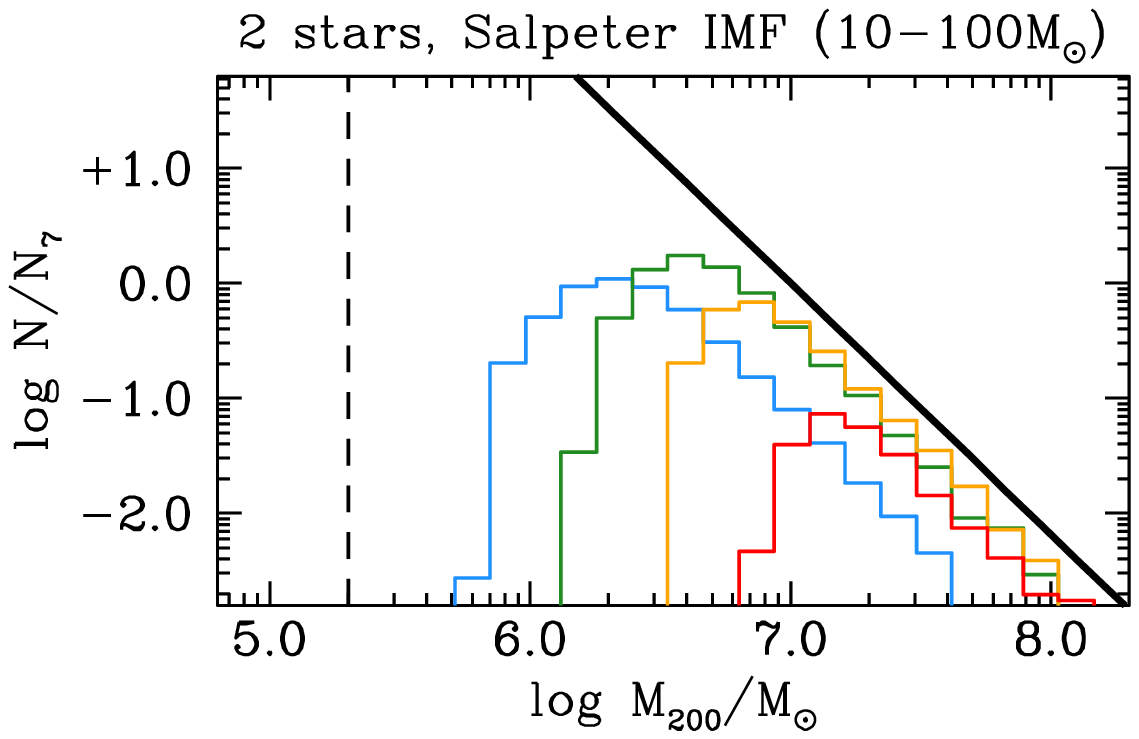}}\\
  \vspace{0.4cm}
  {\includegraphics[angle=0,width=8.0cm]{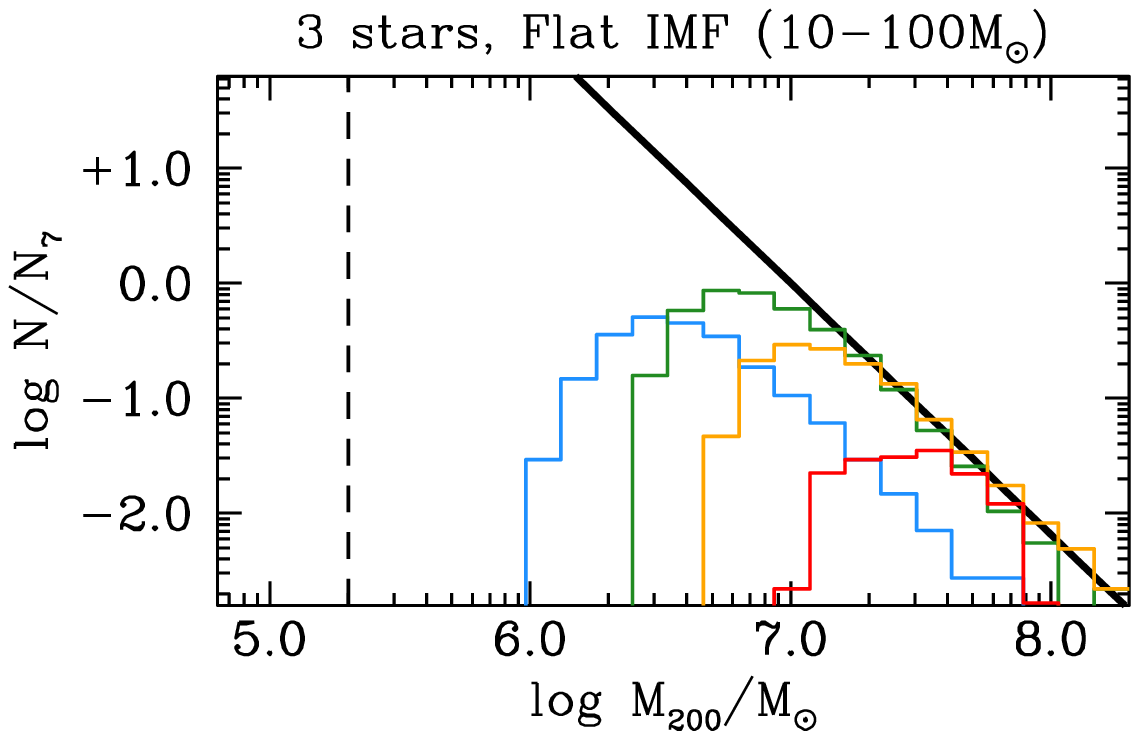}\hspace{0.5cm}
  \includegraphics[angle=0,width=8.0cm]{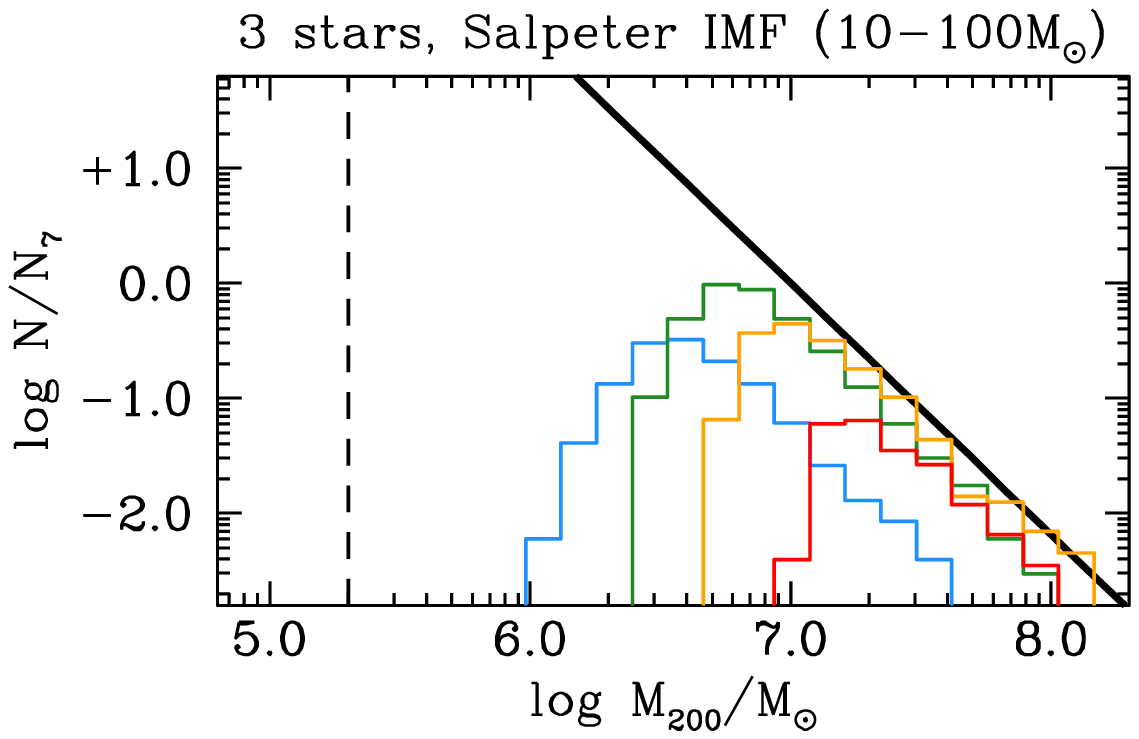}}\\
  \vspace{0.4cm}
  {\includegraphics[angle=0,width=8.0cm]{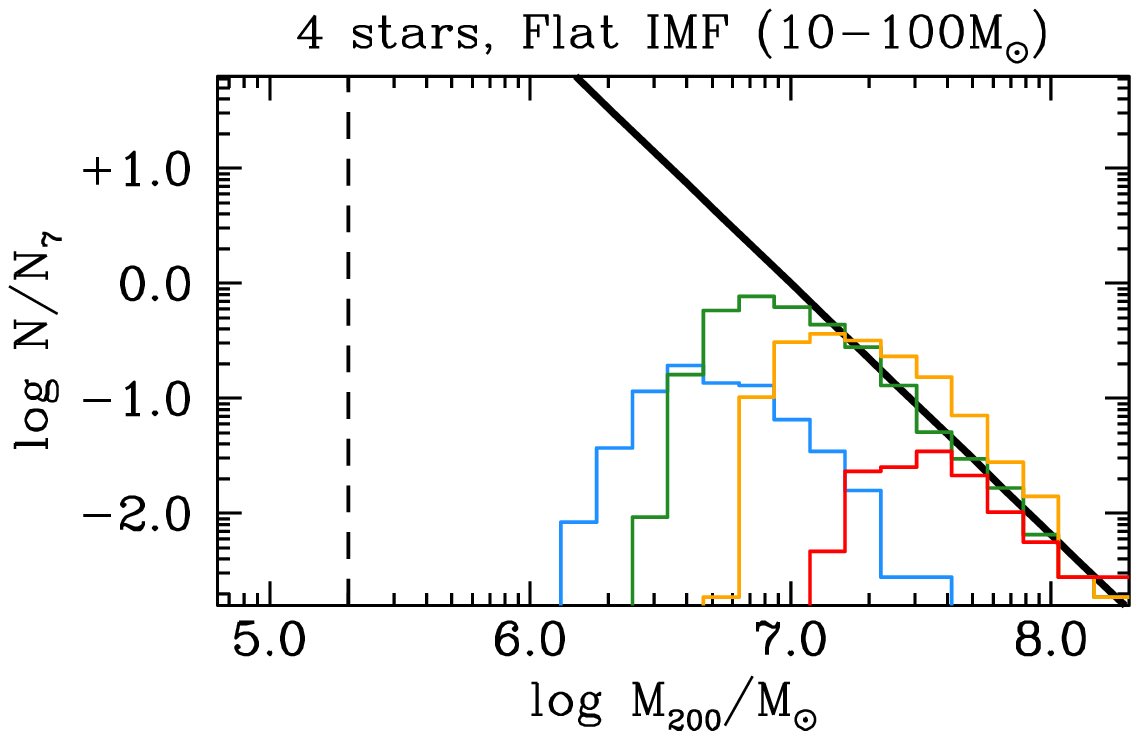}\hspace{0.5cm}
  \includegraphics[angle=0,width=8.0cm]{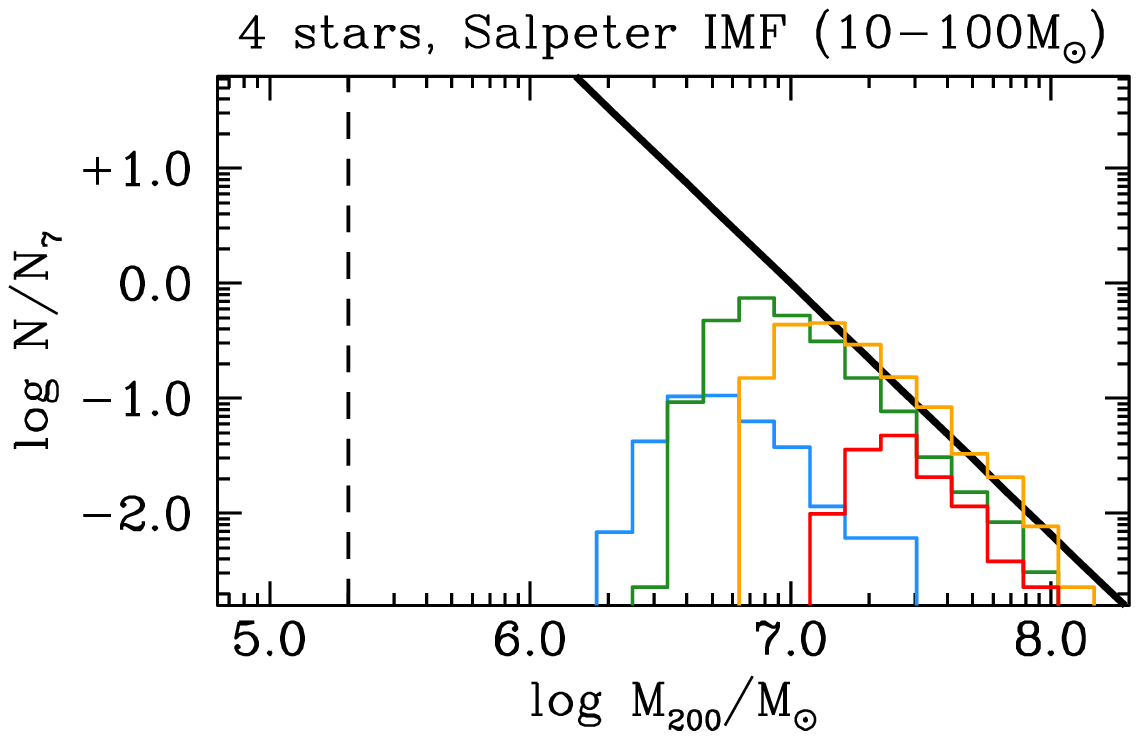}}
  \caption{ 
Distribution of halo masses that retain their gas after hosting a first
generation of stars are shown, color-coded by the average energy released
by the SNe in a given minihalo; blue, green, and orange represent the low-, typical,
and high-energy SNe, while the red curves are for hypernovae. For reference,
the thick black curve illustrates the sampled halo mass function.
The distributions are normalized to the total number of retained models
at $10^{7} {\rm M}_{\odot}$.  Each row shows the four cases considered here,
where a given minihalo hosts one, two, three, or four massive metal-free stars. The left
panels illustrate the retained halo mass function for stars with a flat
IMF, while the right panels are for a Salpeter-like IMF
(both IMFs cover the mass range $10-100\,{\rm M}_{\odot}$).
Minihalos below $2\times10^{5}\,{\rm M}_{\odot}$ (indicated by the vertical dashed
line) are unable to cool sufficiently to form a first stellar generation
\citep{Yos03,OShNor07}.
 }
  \label{fig:massfunc}
\end{figure*}

The working definition of a ``first galaxy'', in its most basic form,
is a dark matter halo that confines a long-lived stellar population
(see e.g. the review by \citealt{BroYos11}). Within our model, the
first galaxies would be those that are the least disrupted by the
SN explosion from the first stars, corresponding to the halos that
are able to retain their gas.

There are two primary mechanisms in our models that influence
the distribution of retained halo masses. Obviously, the most
important factor is the energy that is available to lift the gas
out of the minihalo, which largely determines the shape of the
retained halo mass distribution. However, the form of the
primordial stellar IMF also influences the probability that a
minihalo will retain its gas, since the presupernova gas
distribution strongly depends on stellar masses.
For the example shown in Figure~\ref{fig:densprof},
a $15~{\rm M}_{\odot}$ star, despite having a lower
luminosity than the $40~{\rm M}_{\odot}$ star,
drastically reduces the central gas density of the
minihalo during its relatively longer life. At a fixed
minihalo mass, it is easier for a SN to
evacuate the gas from the minihalos that host
the longest-lived massive stars. Therefore, a
minihalo of a given mass is more likely to retain
its gas if the form of the primordial stellar IMF
is top-heavy.

In Figure~\ref{fig:massfunc}, we present the distribution of halo masses
that retain their gas (color-coded by the average energy of the hosted
SNe). For reference, we also show the halo mass function from which
these models are drawn as a thick black curve. It is immediately
obvious that there is a significant deficit of low-mass halos that
are able to retain their gas. Given that the least disrupted minihalos
are those that are able to \textit{promptly} form a second generation
of stars, our calculations suggest that the second stars were formed
in halos of mass $\sim{\rm a~few}~\times10^{6}~{\rm M}_{\odot}$.

\subsection{Chemistry}
\label{sec:nucleo}

\begin{figure*}
  \centering
  {\includegraphics[angle=0,width=7.3cm]{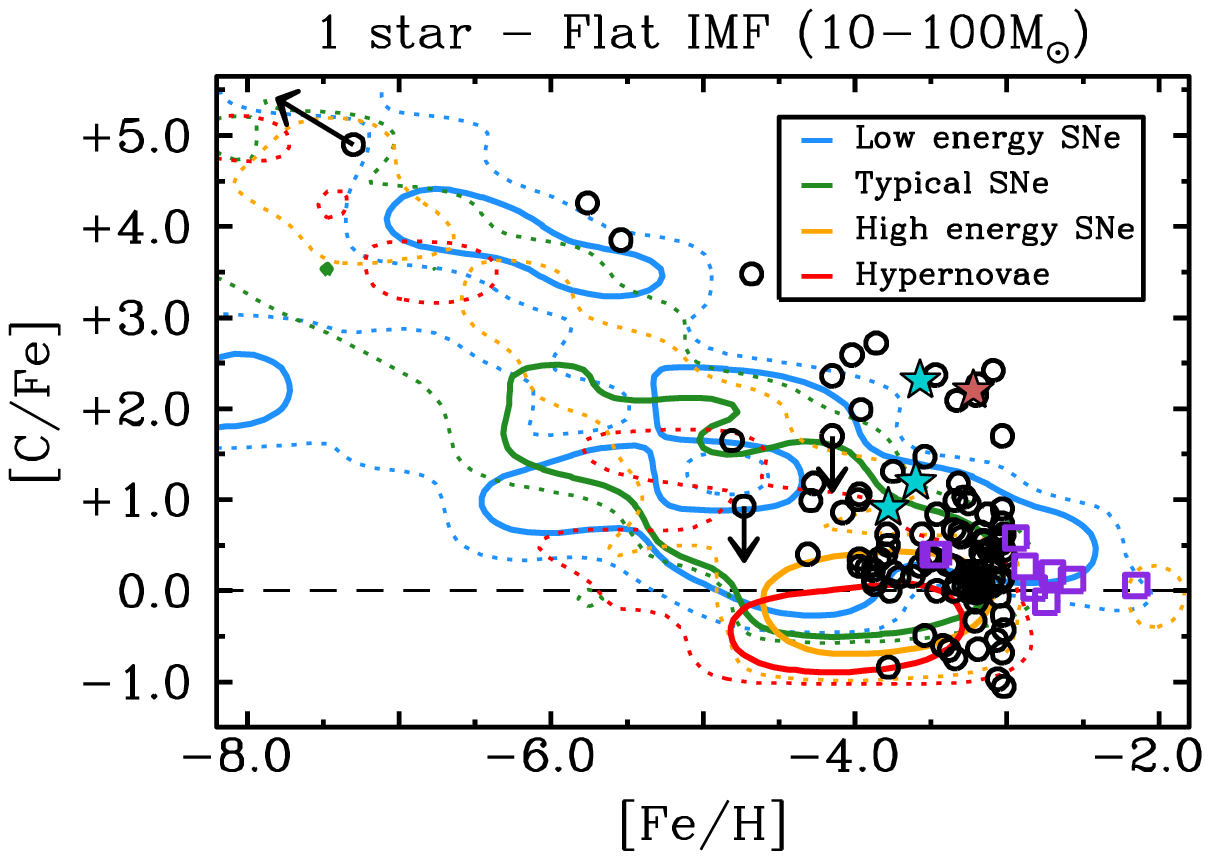}\hspace{1.0cm}
  \includegraphics[angle=0,width=7.3cm]{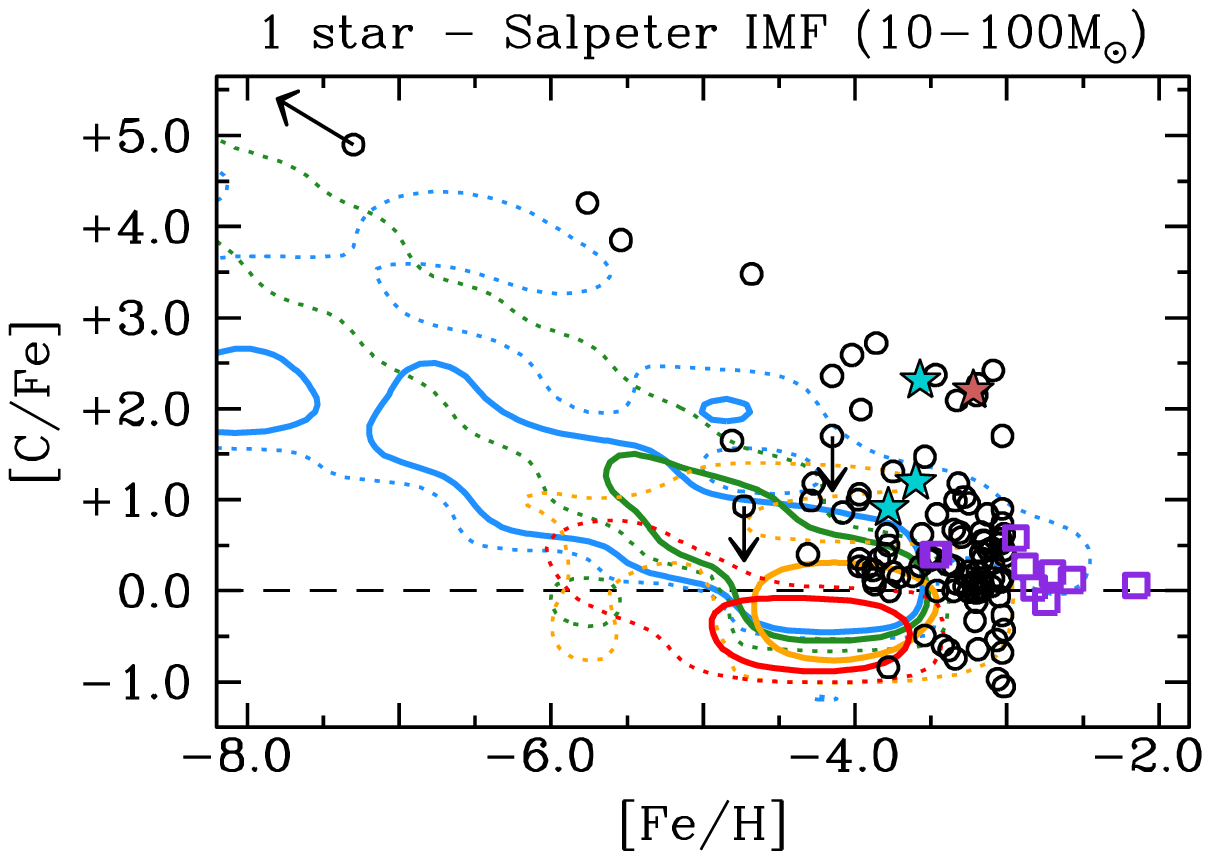}}\\
  \vspace{0.2cm}
  {\includegraphics[angle=0,width=7.3cm]{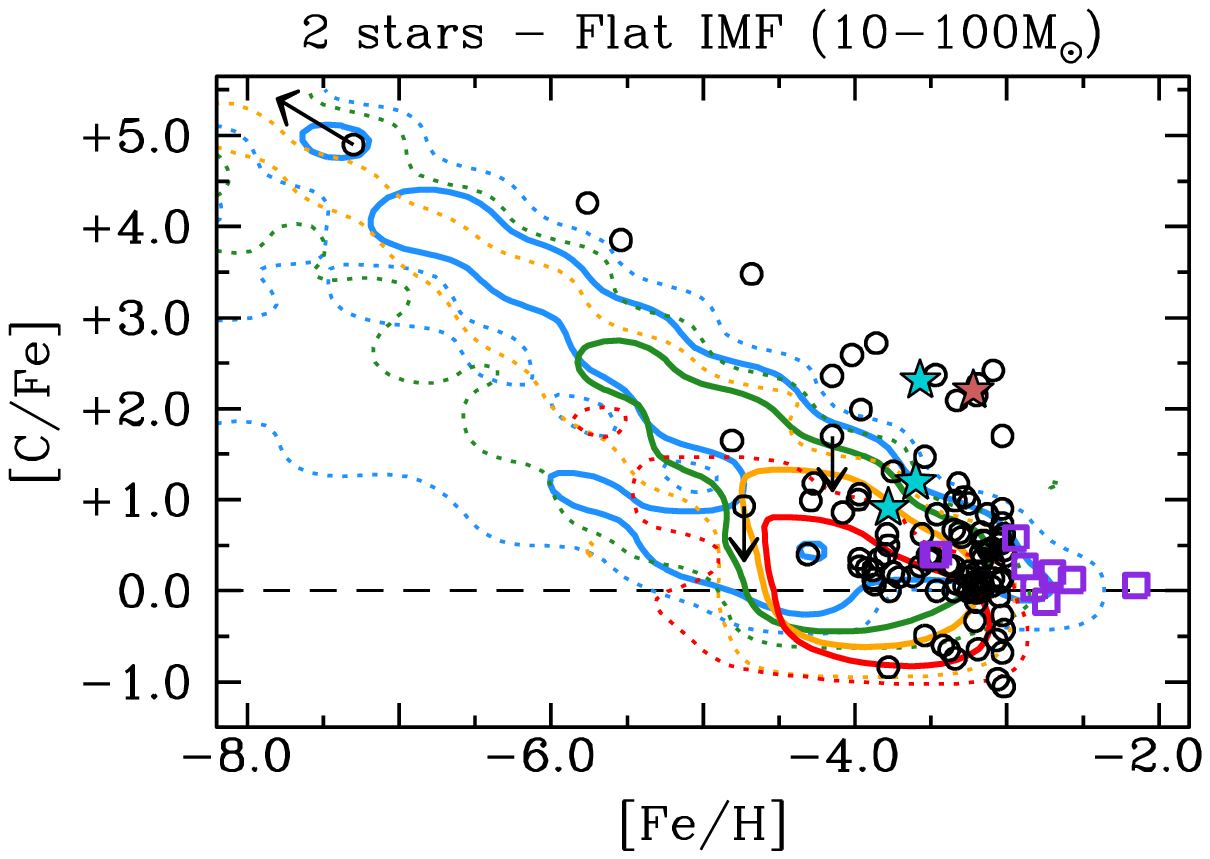}\hspace{1.0cm}
  \includegraphics[angle=0,width=7.3cm]{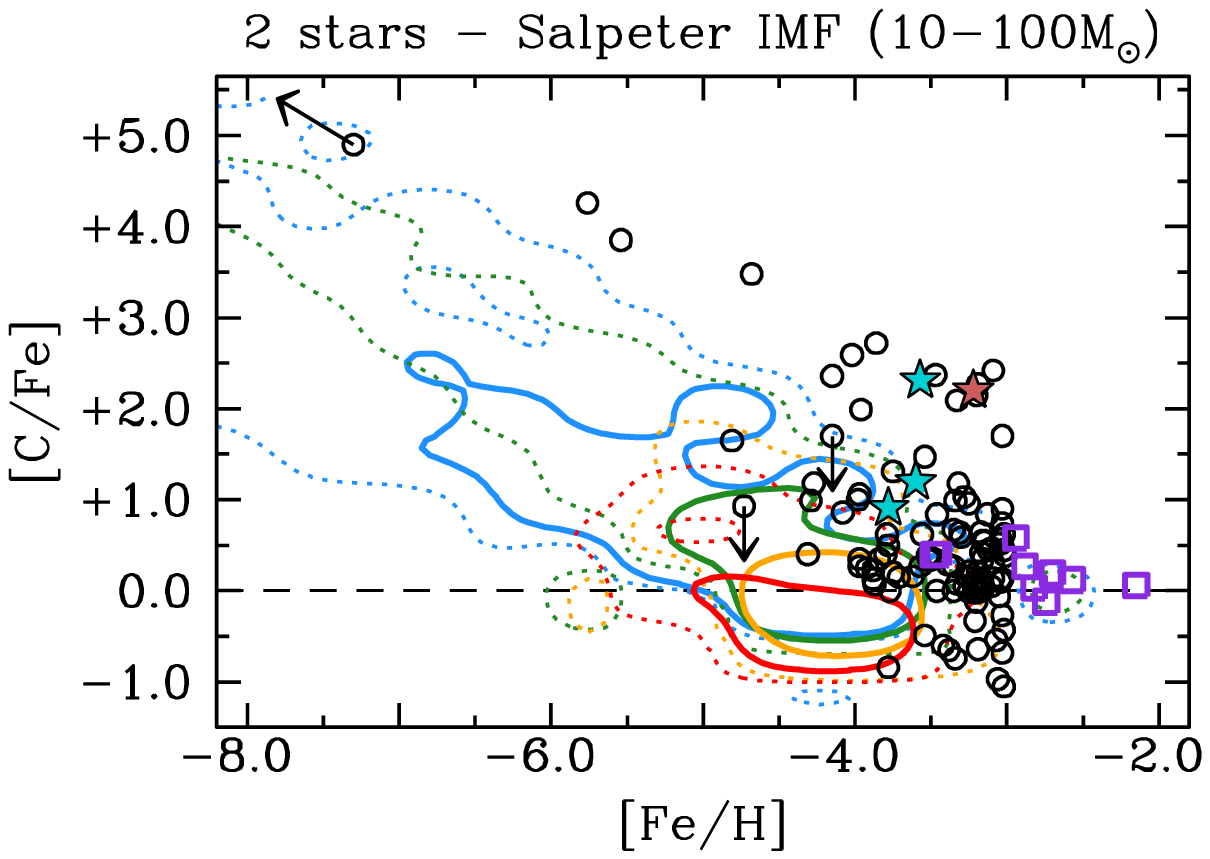}}\\
  \vspace{0.2cm}
  {\includegraphics[angle=0,width=7.3cm]{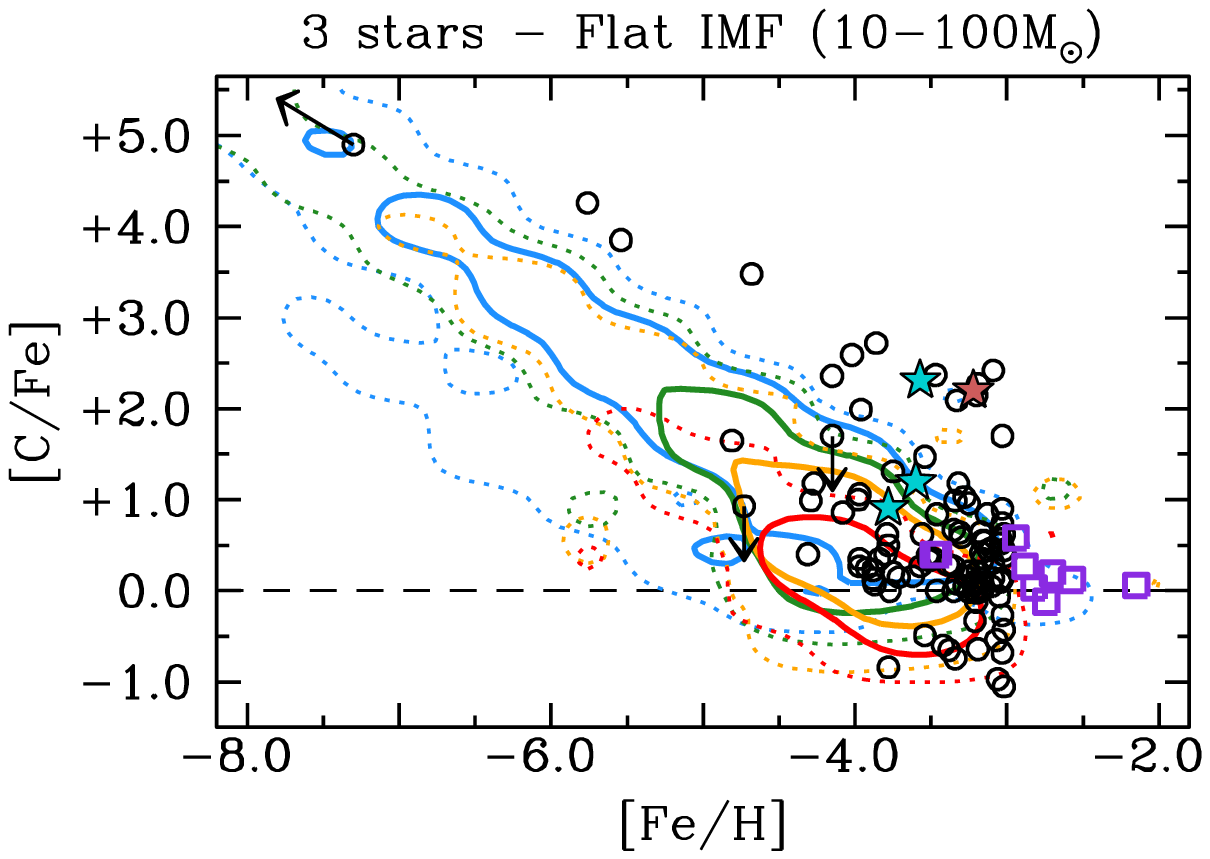}\hspace{1.0cm}
  \includegraphics[angle=0,width=7.3cm]{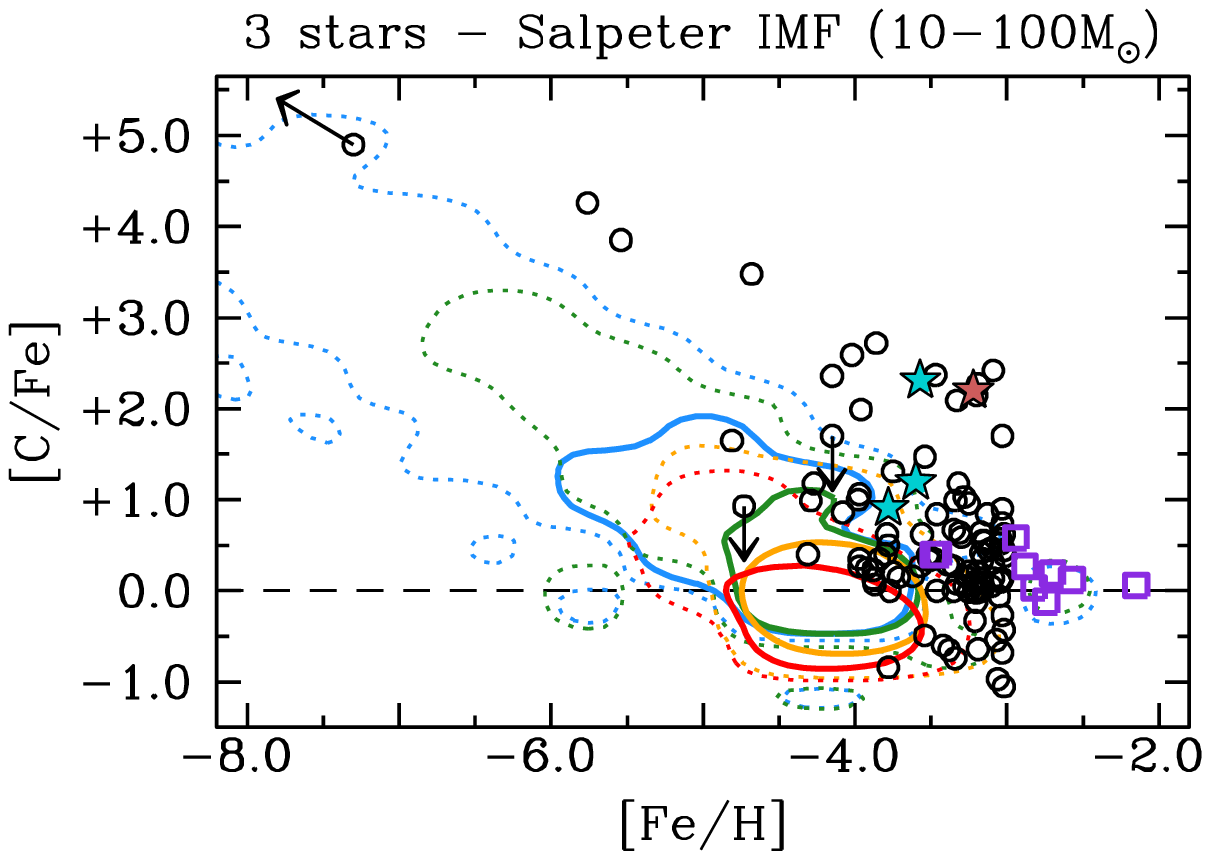}}\\
  \vspace{0.2cm}
  {\includegraphics[angle=0,width=7.3cm]{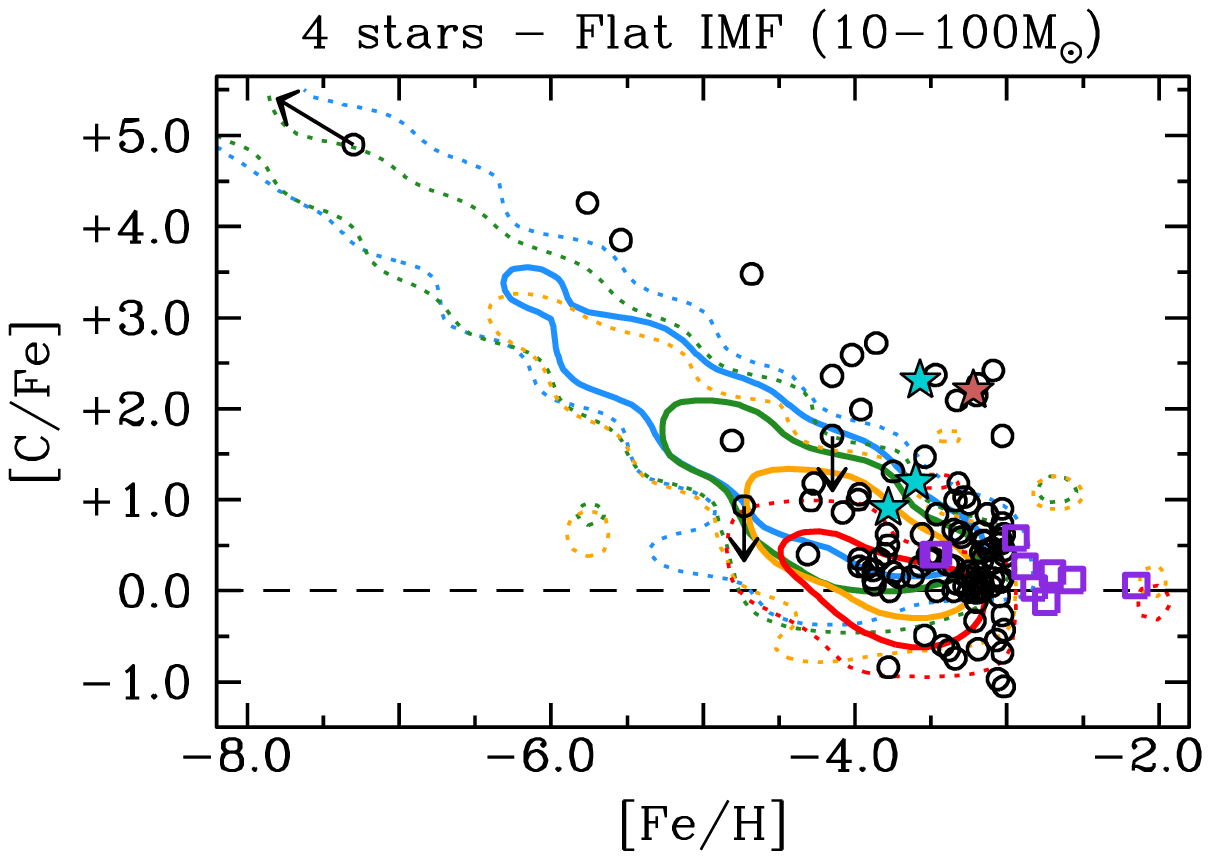}\hspace{1.0cm}
  \includegraphics[angle=0,width=7.3cm]{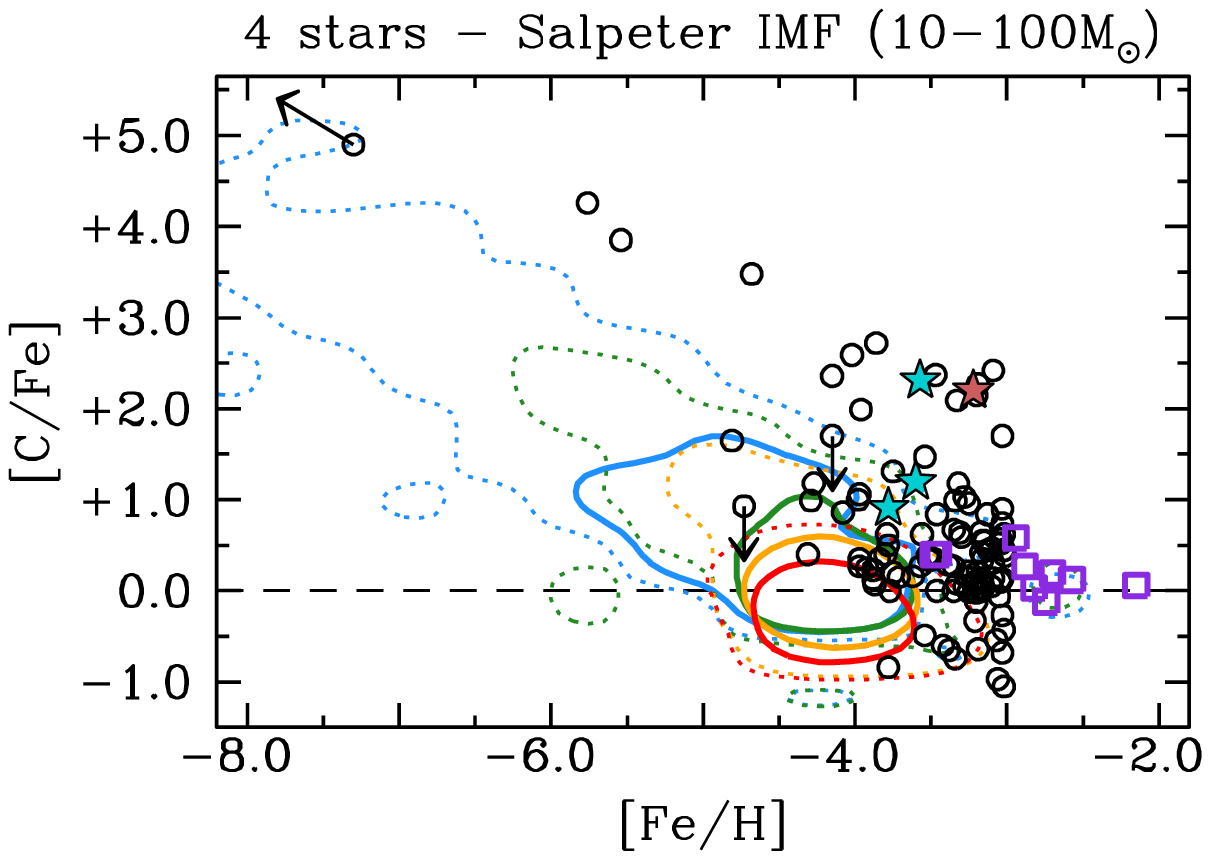}}
  \caption{ 
[C/Fe] vs. [Fe/H] abundances for the halos that retain their gas
after hosting a first generation of stars. The left and right panels display
the distribution of values for a flat and Salpeter-like IMF respectively. From
top to bottom, each panel shows the four cases considered here,
where each minihalo hosts one, two, three, or four massive metal-free stars.
Each set of models is separated based on the average
energy of all SNe for a given minihalo, corresponding to SNe with:
low energy (blue; $0.3-0.75\times10^{51}\,{\rm erg}$),
typical energy (green; $0.75-1.65\times10^{51}\,{\rm erg}$),
high energy (orange; $1.65-4\times10^{51}\,{\rm erg}$) and
hypernovae (red; $4-10\times10^{51}\,{\rm erg}$).
The solid (dotted) curves enclose 68 (95) per cent of the
models for each explosion energy. The black circles are the measurements from
Milky Way stellar halo stars modeled in one dimension and assuming local
thermodynamic equilibrium \citep{Yon13,Aok13,Roe14,Han14}. The two very low-metal
stars reported by \citet{Caf11} and \citet{Kel14}, and one star from \citet{Han14}, are shown
with their derived limits. We also plot four CEMP-no stars that are associated with the
Segue 1 (blue stars; \citealt{Nor10}, \citealt{FreSimKir14}) and
Bo{\" o}tes I (red star; \citealt{Lai11}, \citealt{Gil13})
dwarf spheroidal galaxies. The purple squares are for the
most metal-poor damped Lyman-$\alpha$ systems
\citep{Coo11b,Coo14a,Coo14b}.
The solar level of [C/Fe] is marked by the black horizontal dashed line.
 }
  \label{fig:cfefeh}
\end{figure*}

The chemistry that is born into the second stars is highly sensitive to both
the stellar IMF and the energy released during the SN explosion of a metal-free
star --- these are the two primary mechanisms that also drive
the distribution of retained halo masses. This implies that
there may be a strong selection bias when interpreting the nature
of the first stars by studying their chemical yields in the stellar
atmospheres of the most metal-poor second generation stars in
our Galaxy. The SNe that evacuated the gas from their host minihalo
would have been unable to encode their metal yields in a second generation
of stars. For example, there are no environments currently
known where the distinct chemical signatures
of PISNe (\citealt{HegWoo02})
have been unambiguously detected. This could be a
consequence of an observational survey bias, where the primary
goal is to identify the \textit{most} metal-poor second generation stars;
PISNe are expected to enrich the surrounding medium to a
metallicity of $\sim1/100$ solar \citep{SalSchFer07,KarJohBro08},
which is a considerably higher metallicity than that targeted by
current surveys for metal-poor stars. In addition to this bias, we
suggest that the high energy of their SN explosion would have
evacuated the gas from their host minihalo, resulting in
their chemical signature being rarely incorporated into a
subsequent stellar population (see \citealt{Gre07,Wha08}).
More commonly, one should expect to find the chemical signatures
from the SNe of massive stars that experienced core-collapse, which
typically release more than an order of magnitude less energy than PISNe.
Indeed, the chemical signatures of core-collapse SNe are regularly
observed in the lowest metallicity stars in the halo of our Galaxy
\citep[e.g.][]{McW95,Cay04,Lai08,HegWoo10,Roe14}.

\begin{figure*}
  \centering
  {\includegraphics[angle=0,width=8.3cm]{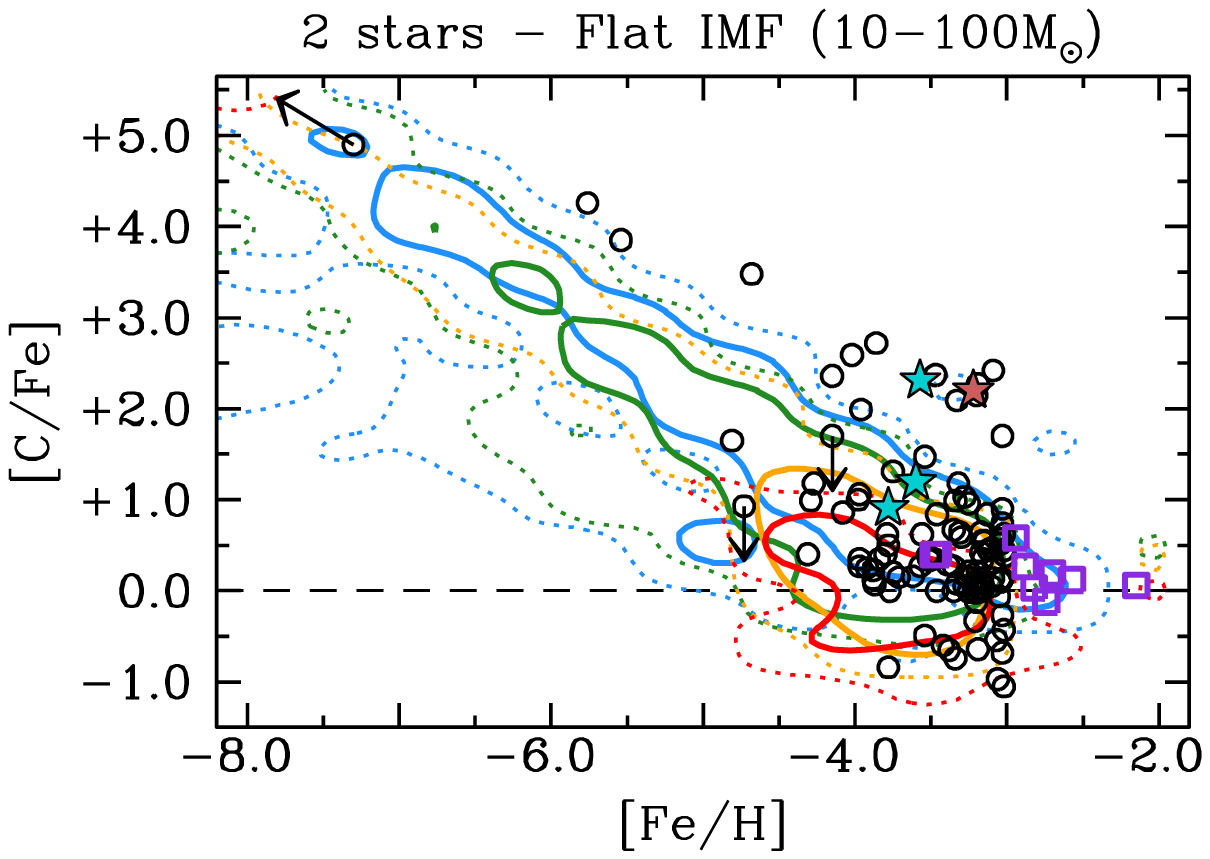}\hspace{1.0cm}
  \includegraphics[angle=0,width=8.3cm]{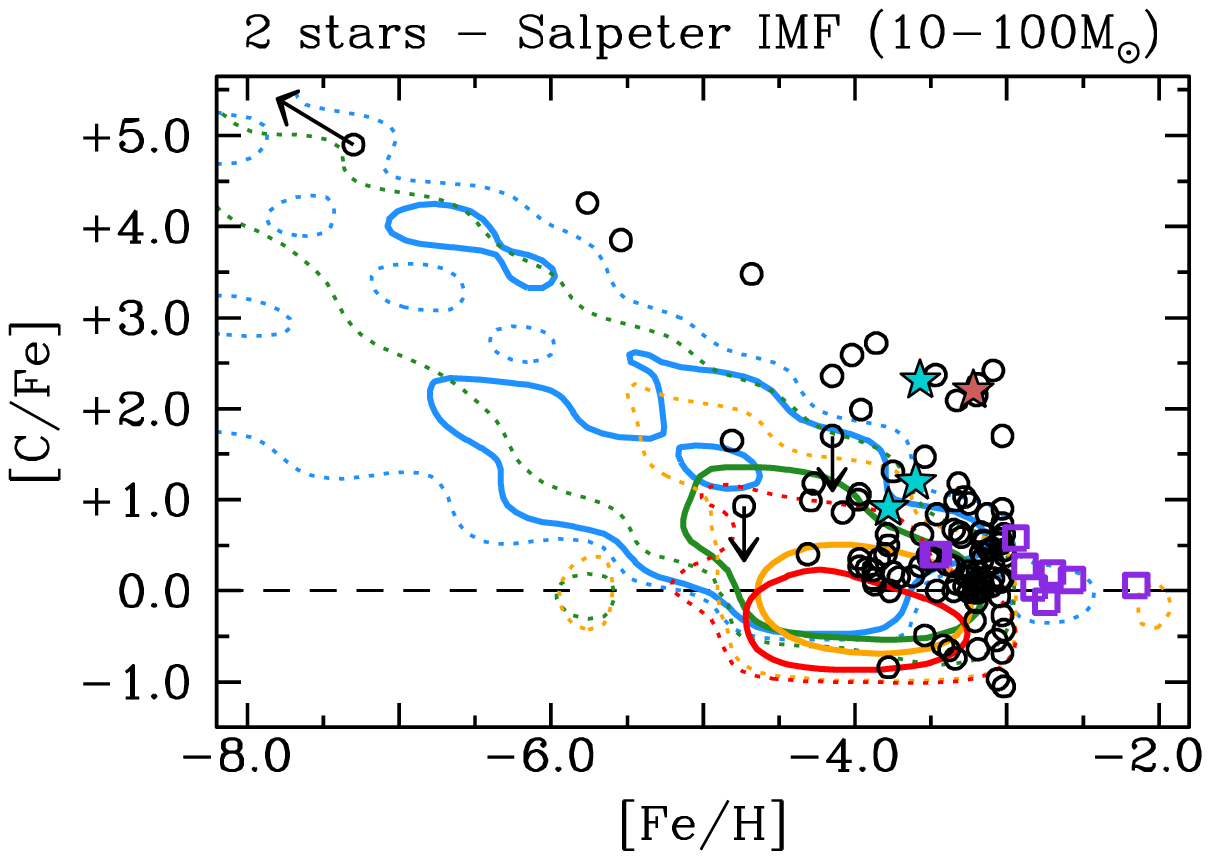}}
  \caption{
Same as Figure~\ref{fig:cfefeh}, however, in this plot we only consider
the minihalos where the SN blastwave does not lift the gas to beyond half the
virial radius. We only illustrate the case where two stars (i.e. a Population III binary)
forms per minihalo; the remaining cases for one, three, or four stars per minihalo
are qualitatively similar to their respective panels in Figure~\ref{fig:cfefeh}.
 }
  \label{fig:halfvir}
\end{figure*}

The strong link between the metal yield, explosion energy, and gas
retention suggests that the second stars should exhibit a strong,
systematic variation in their relative element abundances. Naively,
one should expect that lower atomic number nuclei (e.g. C) are
overabundant relative to the higher atomic number nuclei (e.g. Fe),
since the latter are synthesized in the innermost, tightly bound
core of a massive star. In Figure~\ref{fig:cfefeh}, we show the
distribution of [C/Fe] and [Fe/H] values for the halos that are able
to retain their gas, color-coded by the average energy released
by the SNe in the host minihalo. The solid and dotted contours
enclose 68 and 95 per cent of the models, respectively. We also
overplot the stellar data compiled by \citet{Yon13}
and \citet{Roe14}. We have ignored stars from these samples
with [Ba/Fe]~$> 0.0$, which are probably enriched with the $s$-process.
We supplement these data with three CEMP-no stars from \citet{Aok13},
four stars from \citet{Han14},
and the two metal-poor halo stars reported by \citet{Caf11} and
\citet{Kel14}. We have also included a small handful of stars
from the Milky Way ultra-faint dwarf spheroidal galaxies, including
three stars from Segue 1 \citep{Nor10,FreSimKir14} and
one star from Bo{\" o}tes I \citep{Lai11,Gil13}.

In general, the variation in [C/Fe] as a function of the Fe abundance
for our models show a good agreement with the trend exhibited by
the most metal-poor stars, especially given that there are no free
parameters in our models. In particular, our calculations match the
currently available data for the most Fe deficient CEMP stars, while
simultaneously matching the Leo star \citep{Caf11} and most of the
`carbon-normal' stars just below [Fe/H]~$\simeq-3.0$. All minihalos
are enriched to a total metallicity [Z/H]~$\simeq-3.0$, which is in good
agreement with cosmological hydrodynamic simulations \citep[e.g.][]{Wis12}.
We further note that the most Fe-deficient CEMP stars need not have formed
from the products of a single SN; our models demonstrate that
the most Fe-deficient CEMP stars can also be produced by a small
multiple of first stars.

In Figure~\ref{fig:cfefeh}, there is also a small cluster of $\sim12$ `super' CEMP stars
that are offset from our models toward higher [C/Fe] and higher [Fe/H].
This small handful of stars is clearly distinct from both our model calculations
and from observations of other CEMP stars at a similar metallicity. In addition,
these stars are found in both the Milky Way halo and satellite dwarf galaxies.
One possibility is that these stars may have formed in minihalos that were
enriched by early episodes of AGB nucleosynthesis \citep{SalFer12}.
A similar possibility is that these low-metallicity CEMP-no stars may
have hosted an AGB binary companion, and therefore represent the
`low-$s$' extremely metal-poor counterparts of the CEMP-$s$
population \citep{Mas10}. This scenario is supported by models
of AGB nucleosynthesis at [Fe/H]~$=-3.0$, which predict a [C/Fe]
abundance in the range $+2.0$ to $+3.0$ \citep{CamLat08}. These
models are consistent with the level of enrichment observed for
the super CEMP stars. Dedicated radial velocity monitoring for this
sample of stars is now required to confirm if they are members of
binary systems. At present,
radial velocity variations for 5 of these 12 stars have already been
measured \citep{Sta14}, 3 of which show evidence of
binarity\footnote{The stars that exhibit binarity
are HE\,1150$-$0428, CS\,22957$-$027, and J1422$+$0031. The two stars
with no present evidence for binarity are J1613$+$5309 and Segue 1-7.
The remaining seven stars have undetermined radial velocity variations:
HE\,0233$-$0343, HE\,1012$-$1540, HE\,1310$-$0536, HE\,2139$-$5432,
CS\,22958$-$042, CS\,29498$-$043, and Boo21 \citep{Sta14,Han14}.}.

Alternatively, this handful of CEMP stars could have been born
out of gas that was enriched by metal-free stars exploding as Type-II SNe,
as considered herein, with a boost in the CNO elements provided by
stars that yield no Fe-peak elements. Several models do entertain stellar
yields with no Fe-peak elements, including intermediate mass Population III
stars \citep{Chi01,Her05,CamLat08}, models with rotationally enhanced
mass loss from massive stars \citep{MeyEksMae06,Hir07,Mey10}, and
massive stars that end their lives as a pulsational pair instability SNe
(with progenitor masses in the range $\sim100-140~{\rm M}_{\odot}$,
\citealt{WooHeg14}).

On the other hand, if this cluster of stars does not represent a distinct
population, their disagreement with our models might be a consequence
of our assumption that the metal yields from the first stars are uniformly
mixed with the swept up pristine material. Inefficient mixing has recently
been proposed as a possible explanation for the elevated, constant
[$\alpha$/Fe] ratios seen for several stars over a broad range of metallicity
in the Segue 1 ultra-faint dwarf spheroidal galaxy \citep{FreSimKir14}.
If mixing is considered to be less efficient than allowed by our fiducial
model, the contours in Figure~\ref{fig:cfefeh} would be inflated to account
for regions of relatively higher and lower metallicity. However, inefficient
mixing is unable to explain the absence of stars between the `ordinary'
CEMP stars and the `super' CEMP stars. The details of the
mixing process can only be addressed with more realistic 3D,
cosmological hydrodynamic simulations.

We also note that the [Fe/H] abundance that we derive from our models
depends on the amount of pristine H gas swept up by the blastwave, which is
fixed by the energy delivered by the SN and the potential of the host minihalo.
Our fiducial model (see Section~\ref{sec:retention}) requires that
the gas must not be lifted to beyond the virial radius in order for the minihalo
to retain its gas. If we relax this criterion to allow models where the blastwave
stalls at twice the virial radius, the distributions of the models presented
in Figure~\ref{fig:cfefeh} would be shifted toward lower [Fe/H], since `additional'
pristine gas will be swept up. Similarly, if we
were to enforce a stricter retention criterion, the models shift toward higher [Fe/H].
We explore this possibility in Figure~\ref{fig:halfvir} for the case where two
stars (i.e. a Population III binary) are born per minihalo, and require that each
minihalo is only able to retain its gas if the SN blastwave does not
extend beyond half the virial radius. Enforcing a stricter criterion results
in only a minor shift to the overall distribution of [C/Fe] and [Fe/H]
values. Our retention criterion is therefore relatively insensitive to the
conclusions drawn here regarding the origin of CEMP stars at low
metallicity. As a final note, our models have also assumed that stellar
multiplicity is independent of the minihalo mass. In principle, the
multiplicity of the first stars may depend on the thermal evolution
of the collapsing cloud, which in turn depends on the mass of the
minihalo \citep[e.g.][]{Hir14}. However, this should not significantly
affect the trend of [C/Fe] with [Fe/H]; for a given IMF, our simulation
results exhibit a general similarity in the [C/Fe] -- [Fe/H] plane,
independent of the stellar multiplicity.

\begin{figure*}
  \centering
  {\includegraphics[angle=0,width=7.5cm]{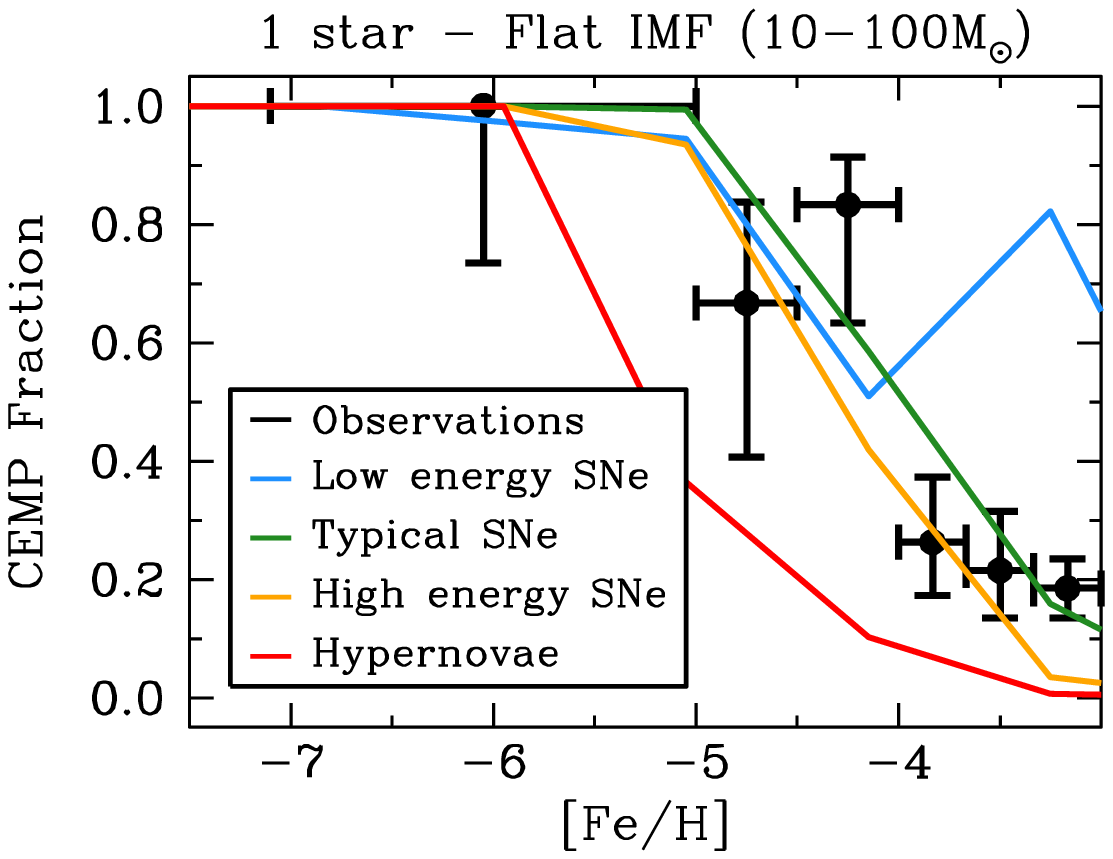}\hspace{1.0cm}
  \includegraphics[angle=0,width=7.5cm]{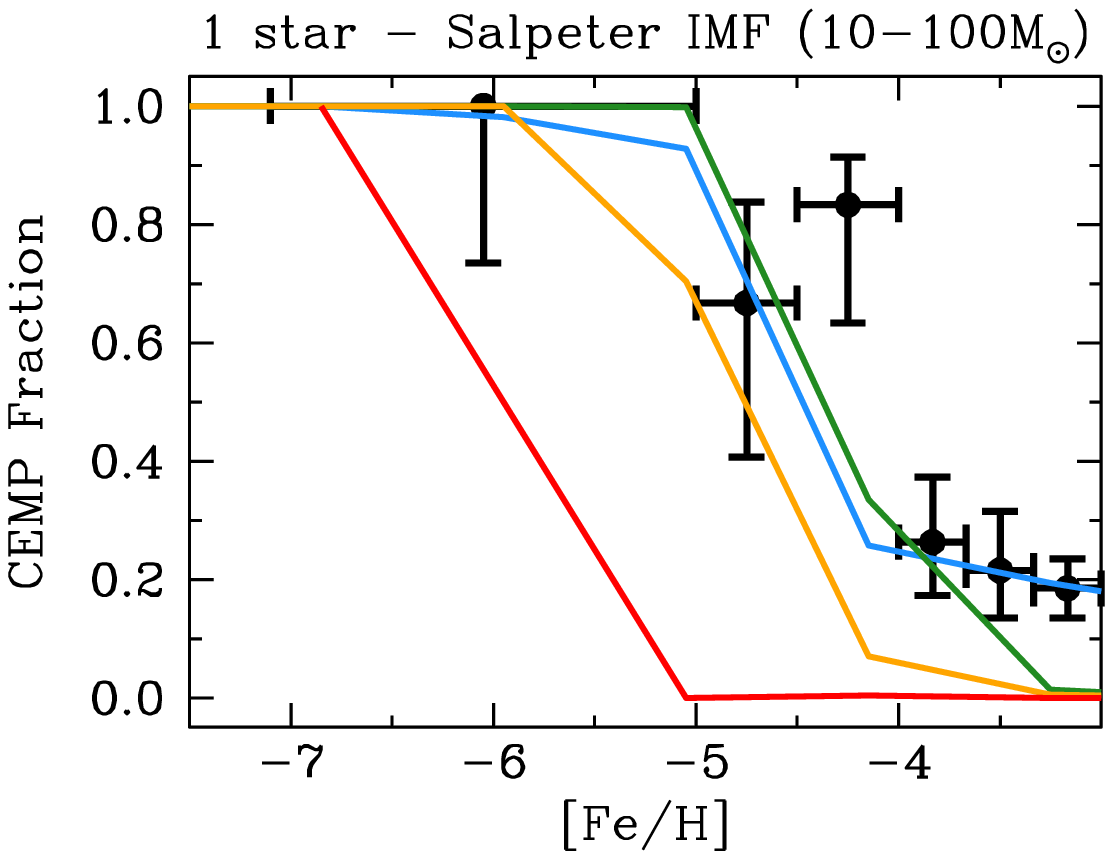}}\\
  \vspace{0.7cm}
  {\includegraphics[angle=0,width=7.5cm]{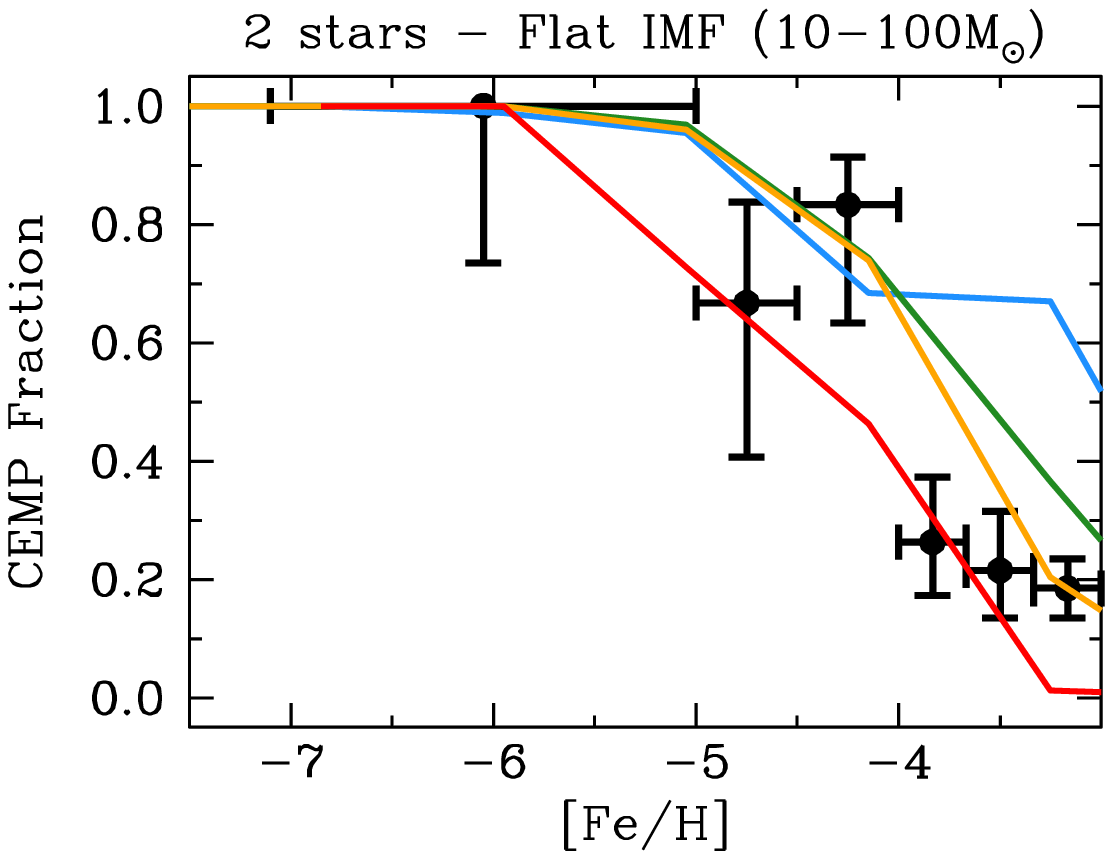}\hspace{1.0cm}
  \includegraphics[angle=0,width=7.5cm]{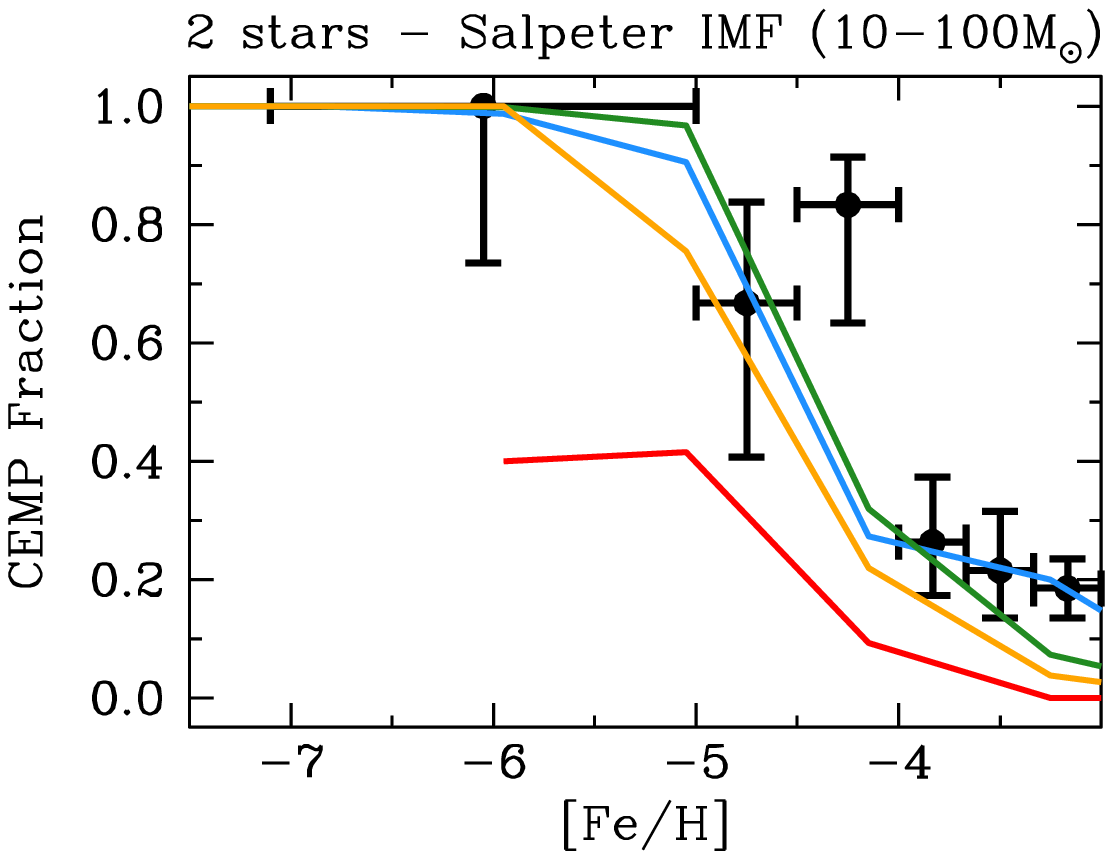}}\\
  \vspace{0.7cm}
  {\includegraphics[angle=0,width=7.5cm]{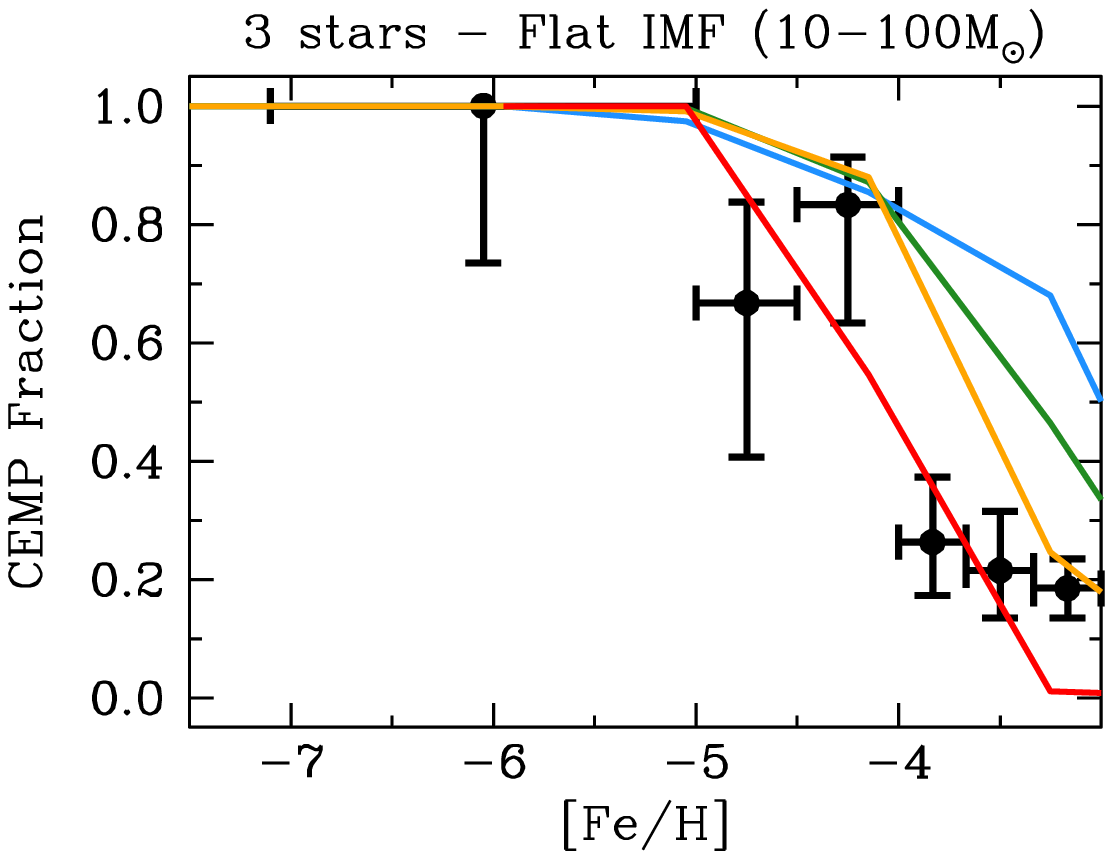}\hspace{1.0cm}
  \includegraphics[angle=0,width=7.5cm]{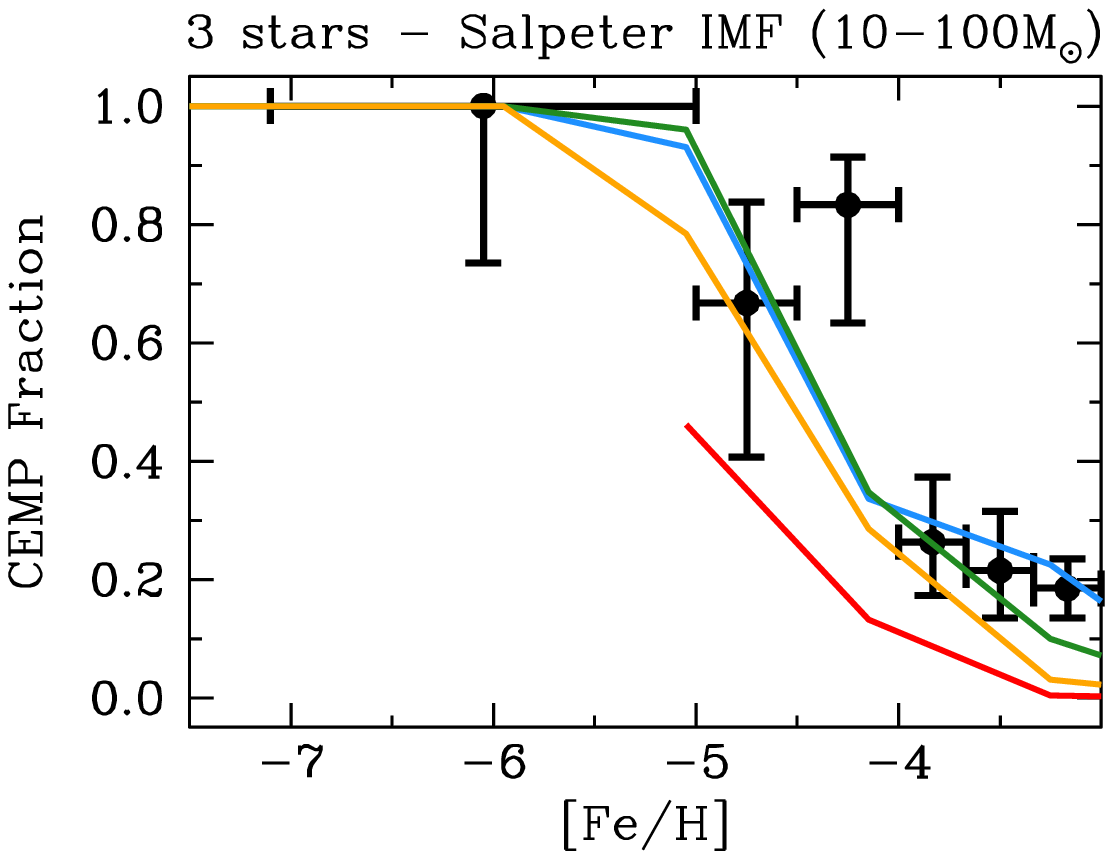}}\\
  \vspace{0.7cm}
  {\includegraphics[angle=0,width=7.5cm]{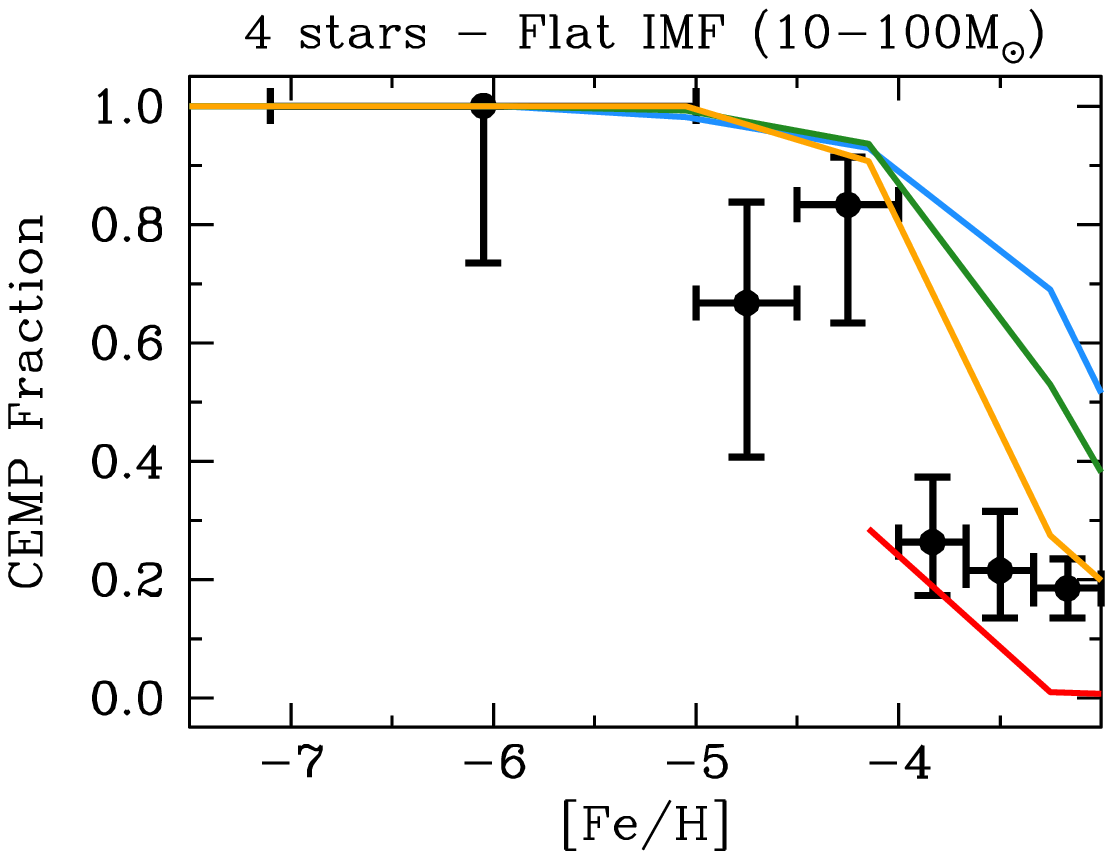}\hspace{1.0cm}
  \includegraphics[angle=0,width=7.5cm]{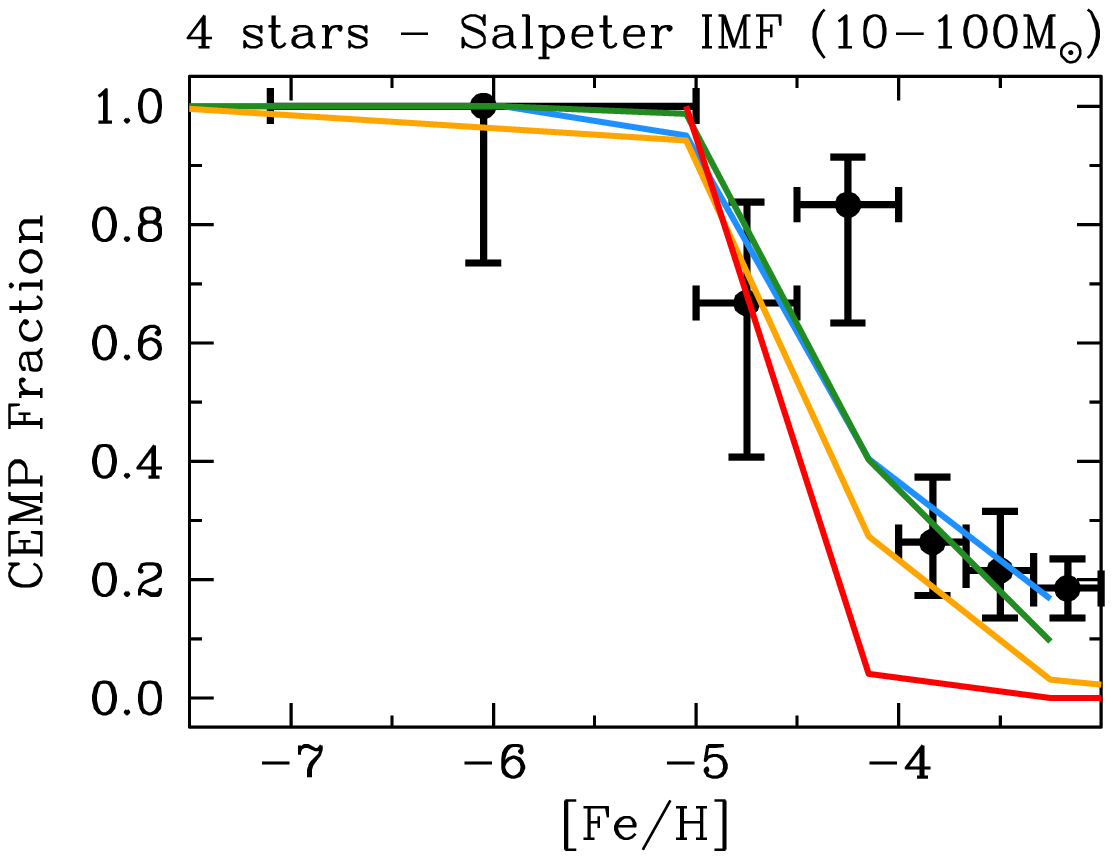}}
  \caption{
Fraction of CEMP stars at a given Fe metallicity is shown for the
halos that retain their gas after hosting a first generation of stars
(same layout and color coding as Figure~\ref{fig:cfefeh}). The black
points are primarily based on the compilations by \citet{Yon13} and
\citet{Roe14}, in addition to three stars from \citep{Aok13}, three stars
from \citet{Han14}, and the two stars reported by \citet{Caf11} and
\citet{Kel14}. We assume the \citet{Aok07} definition of CEMP stars
(i.e. [C/Fe]$\ge+0.7$ for our models).
For a given explosion energy, the model curves are
truncated at the model with the lowest [Fe/H].
 }
  \label{fig:cempfrac}
\end{figure*}

\subsection{The Fraction of Carbon-Enhanced Metal-Poor stars}

The number of stars that exhibit an enhancement in carbon relative to
those stars that do not is known to be much higher at lower Fe abundance
\citep{BeeChr05,Car12,Yon13,Car14}. In the extremely metal-poor regime
([Fe/H]~$\lesssim-3.0$), the dominant class of these carbon-enhanced
stars are those that exhibit normal abundances of their neutron-capture
elements \citep{Aok07,Car14}. In this section, we use our suite of realistic
chemical enrichment models from the first stars to estimate the
fraction of CEMP stars at the lowest metallicities. In doing so,
we assume that all values of the `mixing' parameter
from the \citet{HegWoo10} metal-free nucleosynthesis models are
equally probable. We further assume that each minihalo that is able to retain
its gas will form a second generation of stars. Using the definition
proposed by \citet{Aok07}, we define our models to be carbon-enhanced
when [C/Fe]~$\ge+0.7$. The fraction of our models that produce
carbon-enhanced minihalos as a function of the Fe abundance is
displayed in Figure~\ref{fig:cempfrac} for the two IMFs under consideration.
Using the same criterion, we have calculated the observed fraction
of carbon-enhanced halo stars from the data sample described in
Section~\ref{sec:nucleo}, excluding the four stars that have been
observed in the Segue 1 and Bo{\"o}tes I dwarf spheroidal galaxies.

A general prediction of our models --- independent of the IMF and
gas mixing efficiency --- is that all future stars that are discovered
with an Fe abundance [Fe/H]~$\lesssim-5.0$ should almost certainly
exhibit a carbon enhancement. Before drawing conclusions from the
CEMP fraction at somewhat higher metallicities, we caution that some of
the stars in the observational sample may not have been exclusively
enriched by Population III stars; some of the CEMP stars with
[Fe/H]~$\lesssim-3.0$ may have seen pollution from an extinct
AGB companion star, or otherwise may have condensed from
gas that was also enriched by low-metallicity Population II stars.
These two possibilities would respectively under- or overestimate
the observed CEMP fraction relative to our models. Despite
these potential caveats, our models succeed in producing
a decrease in the CEMP fraction toward higher [Fe/H] that
is in broad agreement with the current observational data.

\section{Summary and Conclusions}
\label{sec:conc}

We have developed a simple formalism with no free parameters to
investigate the detailed chemical enrichment signatures from the
first stars. We propose that there is a strong link between the chemistry
born into the second stars, and the host minihalo's ability to retain
its gas reservoir following the SNe from the first stars.
Qualitatively, massive metal-free stars that produce high-energy
SNe are able to eject significant Fe from their core, but
will also efficiently evacuate the gas from their host minihalo
thereby inhibiting the formation of the second stars. On the other
hand, minihalos that host low-energy SNe are able to
retain their gas reservoir but, as a consequence, yield very little
Fe from the highly bound core of the Pop III star. In both cases, high-energy
and low-energy SNe eject a full complement of
the lower atomic number elements, such as C, which are less
bound to the massive star. The net result is to produce a
significant number of Fe-poor second generation stars that
are highly rich in the low atomic number elements, in accord
with the present observations of the most Fe-poor stars in the
halo of the Milky Way. On the basis of our detailed suite of
calculations that explore an extensive region of parameter
space for the properties of the first stars and their host minihalos,
we draw the following main conclusions.

\smallskip

\noindent ~~(1) In addition to the energy released by the first stars, the
primordial stellar IMF plays an important role in determining whether
or not a minihalo will retain its reservoir of gas, and the resulting
chemistry of the second generation of stars. The longest-lived massive
metal-free stars dramatically reduce the central gas density distribution
during their (relatively) long life. As a consequence, the gas is displaced
to larger radii which indirectly assists the supernova blastwave in
evacuating the gas from the host minihalo.

\smallskip

\noindent ~~(2) The dark matter halos that are the least disrupted
by the first stars will also be the first environments to host a second
stellar generation. In this context, CEMP
stars would have been among the first Population II stars to have
formed in the first galaxies. Our calculations also suggest that the
chemistry that seeded the CEMP stars was in place at very
early times and in relatively low-mass halos
(a~few~$\times10^{6}~{\rm M}_{\odot}$).

\smallskip

\noindent ~~(3) We propose that some carbon-enhanced stars that
contain normal abundances of their neutron-capture elements
(the so-called CEMP-no stars) are the by-product of an intimate
link between the explosions of the first stars and their host
minihalo's ability to retain its gas. Our suite of models are able to
reproduce the general dependence of [C/Fe] with [Fe/H], as well
as the variation in the fraction of CEMP stars relative to non-CEMP
stars as a function of the Fe abundance.

\smallskip

\noindent ~~(4) It is commonly believed that the most Fe-deficient
Galactic halo stars were enriched by just a single metal-free star.
Although our calculations are certainly consistent with this picture,
one cannot rule out the possibility that a small multiple of stars
contributed to the enrichment of these second generation Fe-poor
stars.

\smallskip

\noindent ~~(5) The present lack of chemical evidence for very
massive stars ($\gtrsim140~{\rm M}_{\odot}$) does not suggest
they were uncommon or did not exist. Rather, the large energy
released during their SN explosion could have inhibited
their products from being incorporated into subsequent stellar
populations.

\smallskip

Our models suggest that the first stars commonly enrich some
halos to 1/1000 of solar metallicity, similar to some of
the most metal-poor damped Lyman-$\alpha$ systems already
known at redshift $z\sim3$ \citep{Pet08,Coo11a,Coo11b}.
Therefore, understanding the physical properties of these
near-pristine neutral gas reservoirs, may provide key clues
for understanding the star formation environments of
the second stars \citep{Coo14b}. Finally, the recent discovery
of carbon-enhanced Fe-poor stars in both the Segue 1 and
Bo{\" o}tes I Milky Way dwarf spheroidal galaxies
\citep{Nor10,Lai11,Gil13,FreSimKir14}, adds further
support to the idea that these systems are the surviving
relics of the first galaxies \citep{BovRic09,SalFer09,Mun09}.

\section*{Acknowledgments}
We thank Anna Frebel, Miho Ishigaki, Max Pettini, Stefania
Salvadori, Michele Trenti, and Stan Woosley
for their advice and helpful discussions on various aspects of this paper.
We are grateful to an anonymous referee who made valuable suggestions
that improved this paper.
R.~J.~C. is partially supported by NSF grant AST-1109447.
P.~M. acknowledges support by the NSF through grant
OIA--1124453 and by NASA through grant NNX12AF87G.


\begin{thebibliography}{99}

\bibitem[\protect\citeauthoryear{Abel, Bryan, \& Norman}{2002}]{AbeBryNor02}
Abel T., Bryan G.~L., Norman M.~L., 2002, Sci, 295, 93

\bibitem[\protect\citeauthoryear{Abel, Wise, \& Bryan}{2007}]{AbeWisBry07}
Abel T., Wise J.~H., Bryan G.~L., 2007, ApJ, 659, L87

\bibitem[\protect\citeauthoryear{Alvarez, Bromm, \& Shapiro}{2006}]{AlvBroSha06}
Alvarez M.~A., Bromm V., Shapiro P.~R., 2006, ApJ, 639, 621

\bibitem[\protect\citeauthoryear{Aoki et al.}{2007}]{Aok07}
Aoki W., Beers T.~C., Christlieb N., Norris J.~E., Ryan S.~G., Tsangarides S., 2007, ApJ, 655, 492

\bibitem[\protect\citeauthoryear{Aoki et al.}{2013}]{Aok13}
Aoki W., et al., 2013, AJ, 145, 13

\bibitem[\protect\citeauthoryear{Asplund et al.}{2009}]{Asp09}
Asplund M., Grevesse N., Sauval A.~J., Scott P., 2009, ARA\&A, 47, 481

\bibitem[\protect\citeauthoryear{Baraffe, Heger, \& Woosley}{2001}]{BarHegWoo01}
Baraffe I., Heger A., Woosley S.~E., 2001, ApJ, 550, 890

\bibitem[\protect\citeauthoryear{Beers \& Christlieb}{2005}]{BeeChr05}
Beers T.~C., Christlieb N., 2005, ARA\&A, 43, 531

\bibitem[\protect\citeauthoryear{Bland-Hawthorn, Sutherland, \& Karlsson}{2011}]{BlaSutKar11}
Bland-Hawthorn, J., Sutherland, R., Karlsson, T. 2011, in A Universe of Dwarf Galaxies, ed. M. Koleva, P. Prugniel, I. Vauglin (EAS Publications Series, Vol. 48; Les Ulis: EDP Sciences), 397

\bibitem[\protect\citeauthoryear{Bovill \& Ricotti}{2009}]{BovRic09}
Bovill M.~S., Ricotti M., 2009, ApJ, 693, 1859

\bibitem[\protect\citeauthoryear{Bromm, Coppi, \& Larson}{2002}]{BroCopLar02}
Bromm V., Coppi P.~S., Larson R.~B., 2002, ApJ, 564, 23

\bibitem[\protect\citeauthoryear{Bromm \& Loeb}{2003}]{BroLoe03}
Bromm V., Loeb A., 2003, Nature, 425, 812

\bibitem[\protect\citeauthoryear{Bromm, Yoshida, \& Hernquist}{2003}]{BroYosHer03}
Bromm V., Yoshida N., Hernquist L., 2003, ApJ, 596, L135

\bibitem[\protect\citeauthoryear{Bromm \& Yoshida}{2011}]{BroYos11}
Bromm V., Yoshida N., 2011, ARA\&A, 49, 373

\bibitem[\protect\citeauthoryear{Bromm}{2013}]{Bro13}
Bromm V., 2013, Rep. Prog. Phys., 76, 112901

\bibitem[\protect\citeauthoryear{Caffau et al.}{2011}]{Caf11}
Caffau E., et al., 2011, Nature, 477, 67

\bibitem[\protect\citeauthoryear{Campbell \& Lattanzio}{2008}]{CamLat08}
Campbell S.~W., Lattanzio J.~C., 2008, A\&A, 490, 769

\bibitem[\protect\citeauthoryear{Carollo et al.}{2012}]{Car12}
Carollo D., et al., 2012, ApJ, 744, 195

\bibitem[\protect\citeauthoryear{Carollo et al.}{2014}]{Car14}
Carollo D., Freeman K., Beers T., Placco V., Tumlinson J., Martell S., 2014, arXiv, arXiv:1401.0574

\bibitem[\protect\citeauthoryear{Cayrel et al.}{2004}]{Cay04}
Cayrel R., et al., 2004, A\&A, 416, 1117

\bibitem[\protect\citeauthoryear{Cen}{1992}]{Cen92}
Cen R., 1992, ApJS, 78, 341

\bibitem[\protect\citeauthoryear{Chen et al.}{2014a}]{Che14a}
Chen K.-J., Woosley S., Heger A., Almgren A., Whalen D., 2014a, arXiv, arXiv:1402.4134

\bibitem[\protect\citeauthoryear{Chen et al.}{2014b}]{Che14b}
Chen K.-J., Heger A., Woosley S., Almgren A., Whalen D., 2014b, arXiv, arXiv:1402.5960

\bibitem[\protect\citeauthoryear{Chieffi et al.}{2001}]{Chi01}
Chieffi A., Dom{\'{\i}}nguez I., Limongi M., Straniero O., 2001, ApJ, 554, 1159

\bibitem[\protect\citeauthoryear{Christlieb et al.}{2002}]{Chr02}
Christlieb N., et al., 2002, Nature, 419, 904

\bibitem[\protect\citeauthoryear{Cohen et al.}{2006}]{Coh06}
Cohen J.~G., et al., 2006, AJ, 132, 137

\bibitem[\protect\citeauthoryear{Conroy \& Kratter}{2012}]{ConKra12}
Conroy C., Kratter K.~M., 2012, ApJ, 755, 123

\bibitem[\protect\citeauthoryear{Cooke et al.}{2011a}]{Coo11a}
Cooke R., Pettini M., Steidel C.~C., Rudie G.~C., Jorgenson R.~A., 2011a, MNRAS, 412, 1047

\bibitem[\protect\citeauthoryear{Cooke et al.}{2011b}]{Coo11b}
Cooke R., Pettini M., Steidel C.~C., Rudie G.~C., Nissen P.~E., 2011b, MNRAS, 417, 1534

\bibitem[\protect\citeauthoryear{Cooke, Pettini, \& Murphy}{2012}]{CooPetMur12}
Cooke R., Pettini M., Murphy M.~T., 2012, MNRAS, 425, 347

\bibitem[\protect\citeauthoryear{Cooke et al.}{2014a}]{Coo14a}
Cooke R.~J., Pettini M., Jorgenson R.~A., Murphy M.~T., Steidel C.~C., 2014a, ApJ, 781, 31

\bibitem[\protect\citeauthoryear{Cooke et al.}{2014b}]{Coo14b}
Cooke R.~J., Pettini M., Jorgenson R.~A., 2014b, arXiv:1406.7003

\bibitem[\protect\citeauthoryear{Ellison et al.}{2010}]{Ell10}
Ellison S.~L., Prochaska J.~X., Hennawi J., Lopez S., Usher C., Wolfe A.~M., Russell D.~M., Benn C.~R., 2010, MNRAS, 406, 1435

\bibitem[\protect\citeauthoryear{Frebel et al.}{2005}]{Fre05}
Frebel A., et al., 2005, Nature, 434, 871

\bibitem[\protect\citeauthoryear{Frebel et al.}{2007}]{FreJohBro07}
Frebel A., Christlieb N., Norris J.~E., Thom C., Beers T.~C., Rhee J., 2007, ApJ, 660, L117

\bibitem[\protect\citeauthoryear{Frebel, Simon, \& Kirby}{2014}]{FreSimKir14}
Frebel A., Simon J.~D., Kirby E.~N., 2014, ApJ, 786, 74

\bibitem[\protect\citeauthoryear{Fumagalli, O'Meara, \& Prochaska}{2011}]{FumOMePro11}
Fumagalli M., O'Meara J.~M., Prochaska J.~X., 2011, Sci, 334, 1245

\bibitem[\protect\citeauthoryear{Fryer, Woosley, \& Heger}{2001}]{FryWooHeg01}
Fryer C.~L., Woosley S.~E., Heger A., 2001, ApJ, 550, 372

\bibitem[\protect\citeauthoryear{Gilmore et al.}{2013}]{Gil13}
Gilmore G., Norris J.~E., Monaco L., Yong D., Wyse R.~F.~G., Geisler D., 2013, ApJ, 763, 61

\bibitem[\protect\citeauthoryear{Greif et al.}{2007}]{Gre07}
Greif T.~H., Johnson J.~L., Bromm V., Klessen R.~S., 2007, ApJ, 670, 1

\bibitem[\protect\citeauthoryear{Greif et al.}{2010}]{Gre10}
Greif T.~H., Glover S.~C.~O., Bromm V., Klessen R.~S., 2010, ApJ, 716, 510

\bibitem[\protect\citeauthoryear{Greif et al.}{2011}]{Gre11}
Greif T.~H., Springel V., White S.~D.~M., Glover S.~C.~O., Clark P.~C., Smith R.~J., Klessen R.~S., Bromm V., 2011, ApJ, 737, 75

\bibitem[\protect\citeauthoryear{Greif et al.}{2012}]{Gre12}
Greif T.~H., Bromm V., Clark P.~C., Glover S.~C.~O., Smith R.~J., Klessen R.~S., Yoshida N., Springel V., 2012, MNRAS, 424, 399

\bibitem[\protect\citeauthoryear{Haiman, Thoul, \& Loeb}{1996}]{HaiThoLoe96}
Haiman Z., Thoul A.~A., Loeb A., 1996, ApJ, 464, 523

\bibitem[\protect\citeauthoryear{Hansen, Andersen, \& Nordstr{\" o}m}{2013}]{HanAndNor13}
Hansen T., Andersen J., Nordstr{\"o}m B., 2013, Proceedings of the XII International Symposium on Nuclei in the Cosmos (NIC XII), Proceedings of Science, 146, 193

\bibitem[\protect\citeauthoryear{Hansen et al.}{2014}]{Han14}
Hansen T., et al., 2014, ApJ, 787, 162

\bibitem[\protect\citeauthoryear{Heger \& Woosley}{2002}]{HegWoo02}
Heger A., Woosley S.~E., 2002, ApJ, 567, 532

\bibitem[\protect\citeauthoryear{Heger \& Woosley}{2010}]{HegWoo10}
Heger A., Woosley S.~E., 2010, ApJ, 724, 341

\bibitem[\protect\citeauthoryear{Herwig}{2005}]{Her05}
Herwig F., 2005, ARA\&A, 43, 435

\bibitem[\protect\citeauthoryear{Hirano et al.}{2014}]{Hir14}
Hirano S., Hosokawa T., Yoshida N., Umeda H., Omukai K., Chiaki G., Yorke H.~W., 2014, ApJ, 781, 60

\bibitem[\protect\citeauthoryear{Hirschi}{2007}]{Hir07}
Hirschi R., 2007, A\&A, 461, 571

\bibitem[\protect\citeauthoryear{Ishigaki et al.}{2014}]{Ish14}
Ishigaki M.~N., Tominaga N., Kobayashi C., Nomoto K., 2014, arXiv, arXiv:1404.4817

\bibitem[\protect\citeauthoryear{Ji, Frebel, \& Bromm}{2014}]{JiFreBro14}
Ji A.~P., Frebel A., Bromm V., 2014, ApJ, 782, 95

\bibitem[\protect\citeauthoryear{Karakas \& Lattanzio}{2014}]{KarLat14}
Karakas A.~I., Lattanzio J.~C., 2014, arXiv, arXiv:1405.0062

\bibitem[\protect\citeauthoryear{Karlsson, Johnson, \& Bromm}{2008}]{KarJohBro08}
Karlsson T., Johnson J.~L., Bromm V., 2008, ApJ, 679, 6

\bibitem[\protect\citeauthoryear{Keller et al.}{2014}]{Kel14}
Keller S.~C., et al., 2014, arXiv, arXiv:1402.1517

\bibitem[\protect\citeauthoryear{Klessen, Glover, \& Clark}{2012}]{KleGloCla12}
Klessen R.~S., Glover S.~C.~O., Clark P.~C., 2012, MNRAS, 421, 3217

\bibitem[\protect\citeauthoryear{Komiya et al.}{2010}]{Kom10}
Komiya Y., Habe A., Suda T., Fujimoto M.~Y., 2010, ApJ, 717, 542

\bibitem[\protect\citeauthoryear{Lai et al.}{2008}]{Lai08}
Lai D.~K., Bolte M., Johnson J.~A., Lucatello S., Heger A., Woosley S.~E., 2008, ApJ, 681, 1524

\bibitem[\protect\citeauthoryear{Lai et al.}{2011}]{Lai11}
Lai D.~K., Lee Y.~S., Bolte M., Lucatello S., Beers T.~C., Johnson J.~A., Sivarani T., Rockosi C.~M., 2011, ApJ, 738, 51

\bibitem[\protect\citeauthoryear{Ledoux}{1941}]{Led41}
Ledoux P., 1941, ApJ, 94, 537

\bibitem[\protect\citeauthoryear{Limongi \& Chieffi}{2012}]{LimChi12}
Limongi M., Chieffi A., 2012, ApJS, 199, 38

\bibitem[\protect\citeauthoryear{Lucatello et al.}{2005}]{Luc05}
Lucatello S., Tsangarides S., Beers T.~C., Carretta E., Gratton R.~G., Ryan S.~G., 2005, ApJ, 625, 825

\bibitem[\protect\citeauthoryear{Lucatello et al.}{2006}]{Luc06}
Lucatello S., Beers T.~C., Christlieb N., Barklem P.~S., Rossi S., Marsteller B., Sivarani T., Lee Y.~S., 2006, ApJ, 652, L37

\bibitem[\protect\citeauthoryear{Madau, Ferrara, \& Rees}{2001}]{MadFerRee01}
Madau P., Ferrara A., Rees M.~J., 2001, ApJ, 555, 92

\bibitem[\protect\citeauthoryear{Maio et al.}{2011}]{Mai11}
Maio U., Khochfar S., Johnson J.~L., Ciardi B., 2011, MNRAS, 414, 1145

\bibitem[\protect\citeauthoryear{Marigo, Chiosi, \& Kudritzki}{2003}]{MarChiKud03}
Marigo P., Chiosi C., Kudritzki R.-P., 2003, A\&A, 399, 617

\bibitem[\protect\citeauthoryear{Masseron et al.}{2010}]{Mas10}
Masseron T., Johnson J.~A., Plez B., van Eck S., Primas F., Goriely S., Jorissen A., 2010, A\&A, 509, A93

\bibitem[\protect\citeauthoryear{McClure}{1985}]{McC85}
McClure R.~D., 1985, J. Roy. Astron. Soc. Can., 79, 277

\bibitem[\protect\citeauthoryear{McWilliam et al.}{1995}]{McW95}
McWilliam A., Preston G.~W., Sneden C., Searle L., 1995, AJ, 109, 2757

\bibitem[\protect\citeauthoryear{Meynet, Ekstr{\"o}m, \& Maeder}{2006}]{MeyEksMae06}
Meynet G., Ekstr{\"o}m S., Maeder A., 2006, A\&A, 447, 623

\bibitem[\protect\citeauthoryear{Meynet et al.}{2010}]{Mey10}
Meynet G., Hirschi R., Ekstrom S., Maeder A., Georgy C., Eggenberger P., Chiappini C., 2010, A\&A, 521, A30

\bibitem[\protect\citeauthoryear{Mu{\~n}oz et al.}{2009}]{Mun09}
Mu{\~n}oz J.~A., Madau P., Loeb A., Diemand J., 2009, MNRAS, 400, 1593

\bibitem[\protect\citeauthoryear{Murray, Power, \& Robotham}{2013}]{MurPowRob13}
Murray S., Power C., Robotham A., 2013, arXiv, arXiv:1306.6721

\bibitem[\protect\citeauthoryear{Norris et al.}{2007}]{Nor07}
Norris J.~E., Christlieb N., Korn A.~J., Eriksson K., Bessell M.~S., Beers T.~C., Wisotzki L., Reimers D., 2007, ApJ, 670, 774

\bibitem[\protect\citeauthoryear{Norris et al.}{2010}]{Nor10}
Norris J.~E., Gilmore G., Wyse R.~F.~G., Yong D., Frebel A., 2010, ApJ, 722, L104

\bibitem[\protect\citeauthoryear{Norris et al.}{2013}]{Nor13}
Norris J.~E., et al., 2013, ApJ, 762, 28

\bibitem[\protect\citeauthoryear{O'Shea \& Norman}{2007}]{OShNor07}
O'Shea B.~W., Norman M.~L., 2007, ApJ, 654, 66

\bibitem[\protect\citeauthoryear{Ostriker \& McKee}{1988}]{OstMcK88}
Ostriker J.~P., McKee C.~F., 1988, RvMP, 60, 1

\bibitem[\protect\citeauthoryear{Pettini et al.}{2008}]{Pet08}
Pettini M., Zych B.~J., Steidel C.~C., Chaffee F.~H., 2008, MNRAS, 385, 2011

\bibitem[\protect\citeauthoryear{Planck Collaboration et al.}{2013}]{Efs13}
Planck Collaboration, et al., 2013, arXiv, arXiv:1303.5076

\bibitem[\protect\citeauthoryear{Prada et al.}{2012}]{Pra12}
Prada F., Klypin A.~A., Cuesta A.~J., Betancort-Rijo J.~E., Primack J., 2012, MNRAS, 423, 3018

\bibitem[\protect\citeauthoryear{Reed et al.}{2007}]{Reed07}
Reed D.~S., Bower R., Frenk C.~S., Jenkins A., Theuns T., 2007, MNRAS, 374, 2

\bibitem[\protect\citeauthoryear{Ritter et al.}{2012}]{Rit12}
Ritter J.~S., Safranek-Shrader C., Gnat O., Milosavljevi{\'c} M., Bromm V., 2012, ApJ, 761, 56

\bibitem[\protect\citeauthoryear{Ritter et al.}{2014}]{Rit14}
Ritter, J.~S., Sluder, A., Safranek-Shrader, C., Milosavljevic, M., and Bromm, V., 2014, in preparation

\bibitem[\protect\citeauthoryear{Roederer et al.}{2014}]{Roe14}
Roederer I.~U., Preston G.~W., Thompson I.~B., Shectman S.~A., Sneden C., Burley G.~S., Kelson D.~D., 2014, arXiv, arXiv:1403.6853

\bibitem[\protect\citeauthoryear{Ryan et al.}{2005}]{Rya05}
Ryan S.~G., Aoki W., Norris J.~E., Beers T.~C., 2005, ApJ, 635, 349

\bibitem[\protect\citeauthoryear{Salvadori \& Ferrara}{2009}]{SalFer09}
Salvadori S., Ferrara A., 2009, MNRAS, 395, L6

\bibitem[\protect\citeauthoryear{Salvadori \& Ferrara}{2012}]{SalFer12}
Salvadori S., Ferrara A., 2012, MNRAS, 421, L29

\bibitem[\protect\citeauthoryear{Salvadori, Schneider, \& Ferrara}{2007}]{SalSchFer07}
Salvadori S., Schneider R., Ferrara A., 2007, MNRAS, 381, 647

\bibitem[\protect\citeauthoryear{Scannapieco, Schneider, \& Ferrara}{2003}]{ScaSchFer03}
Scannapieco E., Schneider R., Ferrara A., 2003, ApJ, 589, 35

\bibitem[\protect\citeauthoryear{Schaerer}{2002}]{Sch02}
Schaerer D., 2002, A\&A, 382, 28

\bibitem[\protect\citeauthoryear{Schneider et al.}{2012}]{Sch12}
Schneider R., Omukai K., Limongi M., Ferrara A., Salvaterra R., Chieffi A., Bianchi S., 2012, MNRAS, 423, L60 

\bibitem[\protect\citeauthoryear{Shu et al.}{2002}]{Shu02}
Shu F.~H., Lizano S., Galli D., Cant{\'o} J., Laughlin G., 2002, ApJ, 580, 969

\bibitem[\protect\citeauthoryear{Simcoe et al.}{2012}]{Sim12}
Simcoe R.~A., Sullivan P.~W., Cooksey K.~L., Kao M.~M., Matejek M.~S., Burgasser A.~J., 2012, Nature, 492, 79

\bibitem[\protect\citeauthoryear{Smith et al.}{2014}]{Smi14}
Smith B.~D., Wise J.~H., O'Shea B.~W., and Norman M.~L., in preparation

\bibitem[\protect\citeauthoryear{Stacy \& Bromm}{2013}]{StaBro13}
Stacy A., Bromm V., 2013, MNRAS, 433, 1094

\bibitem[\protect\citeauthoryear{Stacy, Greif, \& Bromm}{2010}]{StaGreBro10}
Stacy A., Greif T.~H., Bromm V., 2010, MNRAS, 403, 45

\bibitem[\protect\citeauthoryear{Starkenburg et al.}{2014}]{Sta14}
Starkenburg E., Shetrone M.~D., McConnachie A.~W., Venn K.~A., 2014, arXiv, arXiv:1404.0385

\bibitem[\protect\citeauthoryear{Tegmark et al.}{1997}]{Teg97}
Tegmark M., Silk J., Rees M.~J., Blanchard A., Abel T., Palla F., 1997, ApJ, 474, 1

\bibitem[\protect\citeauthoryear{Tominaga, Umeda, \& Nomoto}{2007}]{TomUmeNom07}
Tominaga N., Umeda H., Nomoto K., 2007, ApJ, 660, 516

\bibitem[\protect\citeauthoryear{Tumlinson}{2006}]{Tum06}
Tumlinson J., 2006, ApJ, 641, 1

\bibitem[\protect\citeauthoryear{Turk, Abel, \& O'Shea}{2009}]{TurAbeOsh09}
Turk M.~J., Abel T., O'Shea B., 2009, Sci, 325, 601

\bibitem[\protect\citeauthoryear{Umeda \& Nomoto}{2003}]{UmeNom03}
Umeda H., Nomoto K., 2003, Nature, 422, 87

\bibitem[\protect\citeauthoryear{Whalen et al.}{2008}]{Wha08}
Whalen D., van Veelen B., O'Shea B.~W., Norman M.~L., 2008, ApJ, 682, 49

\bibitem[\protect\citeauthoryear{Wise \& Abel}{2008}]{WisAbe08}
Wise J.~H., Abel T., 2008, ApJ, 685, 40

\bibitem[\protect\citeauthoryear{Wise et al.}{2012}]{Wis12}
Wise J.~H., Turk M.~J., Norman M.~L., Abel T., 2012, ApJ, 745, 50

\bibitem[\protect\citeauthoryear{Woosley, Blinnikov, \& Heger}{2007}]{WooBliHeg07}
Woosley S.~E., Blinnikov S., Heger A., 2007, Nature, 450, 390

\bibitem[\protect\citeauthoryear{Woosley \& Heger}{2014}]{WooHeg14}
Woosley S.~E., Heger A., 2014, in Very Massive Stars in the Local Universe, ed. J.~S.~Vink (Springer), in press

\bibitem[\protect\citeauthoryear{Yong et al.}{2013}]{Yon13}
Yong D., et al., 2013, ApJ, 762, 27

\bibitem[\protect\citeauthoryear{Yoshida et al.}{2003}]{Yos03}
Yoshida N., Abel T., Hernquist L., Sugiyama N., 2003, ApJ, 592, 645

\bibitem[\protect\citeauthoryear{Yoshida et al.}{2007}]{Yos07}
Yoshida N., Oh S.~P., Kitayama T., Hernquist L., 2007, ApJ, 663, 687

\end{thebibliography}
\end{document}